\documentclass[a4paper,fleqn]{cas-dc}
\usepackage[numbers]{natbib}
\usepackage{amsmath,amssymb,amsthm}
\usepackage{graphicx}
\usepackage{booktabs}
\usepackage{array}
\usepackage{float}
\usepackage{placeins}
\usepackage{seqsplit}
\newtheorem{proposition}{Proposition}
\newtheorem{corollary}{Corollary}
\newtheorem{definition}{Definition}

\begin{document}
\let\WriteBookmarks\relax
\def\floatpagepagefraction{1}
\def\textpagefraction{.001}

\shorttitle{Passing NPCR without diffusion}
\shortauthors{M. Alfarajat}

\title[mode=title]{Passing NPCR Without Diffusion: Exact Signatures, a
Detection Benchmark and an Audit of Published Image Ciphers}

\author[1]{Mua'ad Alfarajat}[orcid=0000-0003-2489-4857]
\cormark[1]
\ead{m.abufaraj@ju.edu.jo}
\credit{Conceptualization, Methodology, Software, Formal analysis,
        Investigation, Data curation, Validation, Writing -- original draft,
        Writing -- review and editing, Visualization}

\affiliation[1]{organization={Department of Computer Information Systems,
                              School of Information Technology and Systems,
                              The University of Jordan},
            city={Aqaba},
            country={Jordan}}

\cortext[1]{Corresponding author}

\begin{abstract}
NPCR, the share of ciphertext bytes changed by a one-pixel plaintext change,
is reported for nearly every image cipher as evidence of diffusion, with
$99.61\%$ the target. Several mechanisms produce values near it, and the mean
does not say which. We separate three regimes: no propagation, one round of
forward and backward chaining linear over addition or XOR, and independent
ciphertexts. For the second an exact count applies. After a
least-significant-bit change its mean asymptotically matches the
independent-ciphertext reference, while its spread without cancellation is
about five to twenty times larger. In simulation, the one-sided NPCR test of Wu,
Noonan and Agaian flags one-round additive, mixed and S-box chaining in
$2$--$4\%$ of papers; read in both directions, the same five values flag at
least $99.9\%$. Two-sided testing does the detecting; the spread and the
other signatures help diagnose the structure. We then audit
twenty-one published schemes. The count matches a reimplementation of one
published diffusion layer on every trial and accounts for its reported table,
the signatures show two schemes whose printed specifications cannot produce their
reported values, and find that three of our own report plain-versus-cipher
values and fall to two chosen images; ten more change their encryption
material with the plaintext or per encryption, so their NPCR mostly reflects
that change. A
released checker applies the signatures to published tables, showing what a
table measures within stated assumptions; it does not certify security.
\end{abstract}


\maketitle


\section{Introduction}
\label{sec:intro}

Almost every image cipher proposed in the last two decades reports NPCR, the
percentage of ciphertext pixels that change when one plaintext pixel changes.
The target is $99.6094\%$, the value two independent uniform images give, and
a scheme that reports a number close to it is taken to diffuse a local change
across the whole image. Reviewers check the number, authors tabulate it against
earlier schemes, and a value a few hundredths above the target is often read as
an improvement.

This paper shows that the number does not carry that meaning. A cipher can
report $99.61\%$ because it diffuses well, because its keys are derived from a
hash of the image, because the value was computed between the plain image and
its own ciphertext, or because a one-round chaining design, whose average
over positions is the same number at large image sizes, was averaged over
enough trials. The first and the last both exercise the diffusion layer, but
only in the first does the number mean what it is taken to mean; in the other
two the number comes from somewhere else. The diffusion in the title is the
kind the number is taken to certify: a change mixed through the whole
ciphertext as independent ciphertexts would show it. One-round chaining does
propagate a change, but linearly, and it passes without that kind of mixing.

We make this precise by separating three regimes. In regime A a one-pixel
change reaches one ciphertext byte, as it does in any cipher built from a
key-driven permutation and a position-wise mask. In regime B the change is
propagated by one forward and one backward chaining pass, linear over addition
modulo $256$ or over XOR, a recurrent forward--backward chaining structure
used by a subset of the permutation--diffusion ciphers that descend from
Fridrich's design~\cite{Fridrich1998}; Section~\ref{sec:audit} identifies
published schemes that instantiate it. In regime C the two ciphertexts
behave as if independent. The exact NPCR distribution in each regime follows
from a short calculation; the count for regime B rests on elementary difference propagation, and the
results lie in the exact distribution it gives and in what that distribution
shows about published tables. The regime B mean coincides with the regime C mean,
in the limit of large images, when the least significant bit is flipped, so
the mean alone cannot tell the two apart. The spread across trials carries
the information the mean conceals: without a cancellation it is about five
to twenty times the regime C spread at common image sizes, and when a
cancellation lands the overall spread exceeds the regime C spread by more than
two orders of magnitude. Under stated conditions, five reported values are
enough to read it. UACI, under a uniform-byte model, behaves the same way:
its regime B mean is close to the ideal while its spread is about eight
points. None of this certifies a cipher; it tells a reader what a reported
table is and is not evidence of.

\begin{figure*}[!t]\centering
\includegraphics[width=\textwidth]{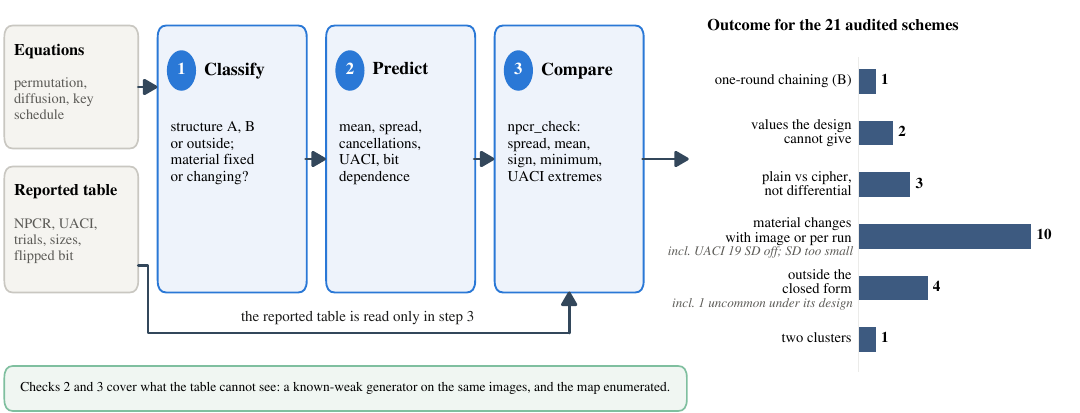}
\caption{Reading a published NPCR table. The printed equations fix the regime
and whether the encryption material is held fixed; the regime fixes the table
that should be reported (Section~\ref{sec:regimes}); the printed table is then
compared with that prediction, a step our checker \texttt{npcr\_check} automates
(Section~\ref{sec:protocol:tool}). The bars give the outcome for the twenty-one
audited schemes of Section~\ref{sec:audit}, a selected set whose counts are
not prevalence estimates. Checks~2 and~3 of
Section~\ref{sec:protocol} test the generator and the map, which the
differential table cannot see.}
\label{fig:workflow}
\end{figure*}

The signatures turn published NPCR tables into evidence that can be checked,
by the three steps of Fig.~\ref{fig:workflow}: classify the structure from
the printed equations, predict the table it should produce, and compare the
prediction with the printed table. We applied them to twenty-one published schemes,
three of them our own.\footnote{Published under the author's former name,
Abu-Faraj.} One scheme from 2026 is in regime B; a reimplementation of its diffusion
layer from the printed equations matches the count exactly, and its reported
NPCR and UACI, mean and spread, are consistent with what the count predicts
for a least-significant-bit change. Two schemes are in regime
A as printed, a one-pixel change reaching at most two bytes, yet report the
mean and spread of regime C. Our own three schemes are in regime A and
reported plain-versus-cipher values; two chosen images decrypt any of their
ciphertexts without the key. Ten schemes derive their keys from the image or add fresh randomness to
each encryption, so their NPCR mostly reflects a change of encryption
material and cannot isolate the diffusion layer. Where they report values, the means sit at
or just above the regime C reference, but that is not the whole table: one
table is unusually tightly grouped under independent trials, and for a 2026
video cipher the UACI extremes printed beside the NPCR lie up to nineteen
standard deviations from the regime C reference. Four fall outside the closed form: two use nonlinear ciphertext
feedback, one chains three rounds of addition and XOR, and one, a batch
cipher, chains once through the whole batch; for the last, both a surrogate
model and a port of the authors' public code produce a table as low as the
printed one in at most $1.7\%$ of runs per batch. One self-adaptive scheme produces a bimodal table whose two clusters the spread
exposes and whose mean, $81.8\%$, would read as weak diffusion.

Each earlier line of work stops short of this. Wu, Noonan and
Agaian~\cite{Wu2011npcr} gave the law of NPCR for independent ciphertexts and
a test against it, but not the values a cipher produces when its ciphertexts
are not independent; their NPCR test rejects only values that are too low,
and it passes one-round chaining. Preishuber et al.~\cite{Preishuber2018}
built insecure ciphers that pass the standard suite, and Ma, Li and
Ou~\cite{Ma2020jisa} observed that insecure ciphers score above $99\%$; both
show that a high NPCR is not evidence, and neither says which structure
produced a given table or how a reader could tell from the table. Teh,
Alawida and Sii~\cite{Teh2020jisa} catalogued the implementation and
reproducibility problems of the field, and Li, Zhang and Xie~\cite{Li2019jisa}
asked for convincing methods of assessing image ciphers in place of the
metrics now in use. For NPCR we offer one: a prediction, from a scheme's
equations, of the table it should report; a check that reads the prediction
off a published table; a benchmark showing which part of that check detects
and which part explains, against the existing test; and an audit that applies
it.

\subsection{Contributions}
\label{sec:intro:contributions}

\begin{enumerate}
\item A diagnostic for NPCR tables: exact counts and a UACI model for three
      regimes, the observation that the mean conceals the regime while the
      spread, under stated assumptions, can point to it, a power calculation stating when five values are
      enough (Section~\ref{sec:regimes}), and a benchmark on nine simulated
      ciphers in which the one-sided NPCR part of the Wu--Noonan--Agaian test
      flags one-round additive, mixed and S-box chaining in $2$ to $4\%$ of
      simulated papers, while a two-sided reading of the same values flags
      nearly all of them; an ablation of nominal $1\%$ tests shows that the
      detection comes from testing both directions and that the spread's role
      is to point to the structure behind a flagged table
      (Section~\ref{sec:protocol:bench}).
\item An audit of twenty-one published schemes, including a structural
      reimplementation of one diffusion layer, a structural model of a batch
      cipher whose reported values are uncommon under two models of its design, a published
      table whose UACI extremes are inconsistent with independent uniform
      ciphertexts although its NPCR is not, and a
      chosen-plaintext break of three of our own (Section~\ref{sec:audit}).
\item A controlled case: $20{,}000$ trials matching the count, a
      three-generator substitution that leaves NPCR and the appearance
      metrics unable to separate a broken generator from ChaCha20, and a
      round-count study on the implementation showing that with extra rounds
      no departure is detected for a low-bit change while the difference map
      stays linear, that a top-bit change leaves the lower seven output bits
      untouched, and that among the round counts tested the mean looks
      reference-like before the dispersion does
      (Sections~\ref{sec:xcm} and~\ref{sec:beyond}).
\item A three-check protocol for authors and reviewers, and the reduced-width
      enumeration that on our own map exposed a degenerate linear core and a
      parity bound that no continuous diagnostic shows
      (Section~\ref{sec:protocol}).
\end{enumerate}

\subsection{What this paper does not claim}
\label{sec:intro:notclaims}

We do not claim that the audited schemes are insecure because of their NPCR
values (the one break we demonstrate, on our own three schemes, follows from
their structure, not from a table), and we do not claim that regime C is sufficient for security; a scheme
whose material changes because it hashes the image has other questions to
answer, whatever its NPCR table shows. Where a
published value disagrees with its printed specification we report the
disagreement and do not guess at its cause. The cipher of Section~\ref{sec:xcm}
is a measurement instrument, not a proposal; we give no confidentiality proof
for it and it is far slower than standardised authenticated encryption.

Section~\ref{sec:background} gives the background. Sections~\ref{sec:regimes}
and~\ref{sec:audit} contain the main results, Sections~\ref{sec:xcm}
and~\ref{sec:beyond} the controlled experiments, Section~\ref{sec:protocol}
the protocol, and Section~\ref{sec:discussion} the limitations.


\section{Background and related work}
\label{sec:background}

\subsection{NPCR and its reference values}

Let $C_1$ and $C_2$ be the ciphertexts of two plaintexts that differ in one
pixel, encrypted under the same key. NPCR is the fraction of positions at
which $C_1$ and $C_2$ differ, and UACI the mean absolute difference divided by
$255$. Wu, Noonan and Agaian~\cite{Wu2011npcr} derived the mean and variance
of NPCR under the hypothesis that the two ciphertexts are independent and
uniform, and turned them into tests with critical values that depend on the
image size. Their mean for $8$-bit images is the familiar $99.6094\%$. The NPCR test
rejects only values below its critical value, so it detects diffusion that is
too weak and has nothing to say about values that sit above the mean; their
UACI test is two-sided. Both apply to one measurement, not to the distribution
of values a paper reports.

Zhang~\cite{Zhang2021ins} took the statistics further. He derived the expected
values and variances of NPCR, UACI and his blocked variant BACI under the same
ideal model, built acceptance ranges for them from the central limit theorem,
and proposed judging a cryptosystem over repeated tests, stating the
criterion for BACI: the share of tests that pass, or how closely averaged
values agree. That work answers how much
a sensitivity index should vary when the ciphertexts behave ideally, and how
to use many tests rather than one; it also checks five published
cryptosystems against those ranges. It gives no distribution for the indices
when the ciphertexts do not behave ideally, and that is where our additions
sit: the exact distribution under one specified non-ideal structure,
one-round chaining, with its cancellations and its dependence on the changed
bit; the separation of
fixed-material from changing-material experiments, which decides what any of
these tests is measuring; and the use of these predictions to read published
tables, including tables whose values pass every acceptance range because the
material, not the diffusion layer, produced them.

The definition leaves two things to the author, and both matter. The first is
which bit or which value is changed. The second is what the second ciphertext
is. Most papers state that one pixel was changed; few state which bit, and some
compute the statistic between the plain image and its ciphertext instead of
between two ciphertexts. Section~\ref{sec:regimes} shows that each choice
changes the result in a predictable way.

\subsection{The permutation--diffusion architecture and its critics}

Fridrich~\cite{Fridrich1998} introduced the architecture that many image
ciphers still follow in some form: a key-driven permutation of pixel positions, then a diffusion
pass in which each output byte depends on the current byte, a keystream byte
and the previous output. Regimes A and B of Section~\ref{sec:regimes} are
this architecture without and with the chaining term. The standard evaluation
of such a cipher (histogram, entropy, adjacent-pixel correlation, NPCR, UACI
and key sensitivity) has been questioned for as long as it has been used.
Alvarez and Li~\cite{AlvarezLi2006} listed requirements for chaos-based
cryptosystems and observed that a positive Lyapunov exponent, a large nominal
key space and noise-like output are neither necessary nor sufficient for
security. Preishuber et al.~\cite{Preishuber2018} showed that the usual test
suite passes constructions with no security value and concluded that the
measurements do not discriminate. Li, Zhang and Xie~\cite{Li2019jisa}, reviewing
the image-cipher designs and cryptanalyses of 2018, found the common metrics widely misused
and called for convincing methods of assessment; Ma, Li and
Ou~\cite{Ma2020jisa}, after breaking a block image cipher with five chosen
images, judged the NPCR criterion not credible, since ciphers known to be
insecure score above $99\%$ on it. Teh, Alawida and Sii~\cite{Teh2020jisa}
revisited the implementation, efficiency and reproducibility problems of the
field and found them largely unaddressed in recent work; Abba, Teh and
Alawida~\cite{Abba2024keyspace} show that the nominal key spaces reported
for such ciphers overstate the effective ones, and Everett~\cite{Everett2024csf}
examines from the dynamical-systems side what a discretised chaotic map can
and cannot provide to a cipher. That a chaotic map iterated on a finite grid
is a different object from the map itself is older than these ciphers.
Lanford~\cite{Lanford1998} compared the orbit structure of a discretised
expanding map with that of a random mapping on the same set; Diamond, Kloeden,
Pokrovskii and co-authors showed that discretisation can collapse chaotic
behaviour onto trivial dynamics or short
cycles~\cite{DiamondSuzuki1996,Diamond1997collapse}; and Guih\'eneuf proved
that for generic conservative systems a single discretisation reflects the
grid more than the map~\cite{Guiheneuf2015homeo}, that computed orbit
segments cannot single out the physical measure~\cite{Guiheneuf2018measures},
and that for generic smooth maps, expanding circle maps among them, the
fraction of grid points lying on cycles tends to zero as the grid is
refined~\cite{Guiheneuf2019recurrence}; repeated discretised rotations of a
pixel grid likewise lose image information~\cite{Guiheneuf2016isometries}. On the
cryptographic side, Li, Chen and Mou~\cite{LiChenMou2005} showed that digital
chaotic maps degrade under finite precision, Persohn and
Povinelli~\cite{PersohnPovinelli2012} found a logistic-map generator
exponentially worse than conventional ones in effective bit length, Galias
enumerated the cycles of the logistic map in single and double
precision~\cite{Galias2021logistic}, and later work analyses implemented maps
through their state-mapping
networks~\cite{LiFengLi2019smn,FanDing2023enumeration,LiTan2024baker}.

Preishuber et al.\ made their case by counterexample: they built ciphers that
pass the suite and are plainly insecure. Wu, Noonan and Agaian supplied the
reference law and a test for one hypothesis, independence. What neither
gives is how to read a value that a real scheme reports when that hypothesis
is false. Table~\ref{tab:priorwork} states what this paper adds to each line
of prior work: not the general inadequacy of the metrics, which is
established, but the distribution of one metric under a common
structured alternative and an audit of published tables against it.

\begin{table}[!t]\centering
\caption{What this paper adds to each line of prior work.}
\label{tab:priorwork}\small
\begin{tabular}{@{}>{\raggedright\arraybackslash}p{2.1cm}>{\raggedright\arraybackslash}p{2.6cm}>{\raggedright\arraybackslash}p{2.8cm}@{}}\toprule
Prior work & Existing result & Added here\\ \midrule
Reference tests~\cite{Wu2011npcr,Zhang2021ins} & NPCR/UACI moments and acceptance ranges under independence; per-value and repeated-test criteria & Distribution under one-round chaining; cancellation and bit dependence; fixed versus changing material; benchmark of the tests (Section~\ref{sec:protocol:bench})\\
Evaluation critiques~\cite{AlvarezLi2006,Preishuber2018,Li2019jisa,Ma2020jisa,Teh2020jisa} & Weak schemes pass the suite; metrics misused & Equation-based prediction of reported values; audit of 21 tables; a checker for reviewers\\
Permutation--diffusion cryptanalysis~\cite{Solak2010,Xie2017,LiLo2011,ChenChenZhou2021universal,ZhangChenChen2022bilateral,Singh2024jisa,ZhouYu2026jisa} & Equivalent masks, permutations, affine maps; breaks of published schemes & Consequence for what the differential test reports, at any key\\
Digital chaos~\cite{Lanford1998,Guiheneuf2015homeo,LiChenMou2005,LiFengLi2019smn,Everett2024csf} & Finite-precision degradation & Structural bounds for one map, as a control\\
\bottomrule\end{tabular}\end{table}

\subsection{Cryptanalysis of permutation--diffusion ciphers}

A second body of work breaks specific designs. Permutation-only ciphers, our
regime A, fall to a small number of chosen or known plaintexts; Li and
Lo~\cite{LiLo2011} give the optimal count, and the two-image break of
Section~\ref{sec:audit} is a special case of their result. Fridrich's own
scheme was broken by Solak et al.~\cite{Solak2010} and, with a sharper attack,
by Xie et al.~\cite{Xie2017}; both exploit the way a single difference
propagates through one diffusion pass, which is the same propagation the
regime B count describes. Chen, Chen and Zhou give a chosen-ciphertext attack
on a family of permutation--substitution schemes with modular
arithmetic~\cite{ChenChenZhou2021universal} and a basis attack on XOR-based
structures~\cite{ChenChenZhou2021psn}; Zhang, Chen and Chen reduce a forward
and backward diffusion to simpler equivalent
operations~\cite{ZhangChenChen2022bilateral}; Li et al.\ break a design whose
diffusion mixes modular addition and XOR~\cite{Li2013modxor}. Recent work
continues the pattern. Singh et al.~\cite{Singh2024jisa} recovered the content
of a multi-image cipher~\cite{Patro2020jisa} from its ciphertext alone by
undoing its feed-forward XOR, and noted that its statistical results had been
in the standard range; Wong et al.~\cite{Wong2024jisa} broke a scheme whose
key depends on pixel sums by choosing images with equal sums; and Zhou and
Yu~\cite{ZhouYu2026jisa} recovered the permutations of a two-round
additive-chaining cipher from chosen-plaintext differences, observing that NPCR
and UACI meeting their targets does not make an algorithm safe. These attacks
recover secret or equivalent encryption information. The regime B count rests on the same linearity but answers a
different question: it predicts the value the standard differential test
reports, for any key, which is what lets a reader check a published number.

\subsection{Plaintext-dependent keys and nonces}

Many recent schemes derive their chaotic parameters from a hash or a sum of
the plain image. The motivation is sound: in a reused permutation-and-mask
construction, chosen plaintexts can reveal an equivalent mask and permutation,
as our own earlier schemes illustrate in Section~\ref{sec:audit}. The consequence for NPCR is less often
noticed. A one-pixel change alters the hash, the hash replaces the keystream,
and the two ciphertexts are produced under different keystreams before any
diffusion takes place. The value reported then mostly reflects that
replacement of material; whether the result behaves as independent ciphertexts is a property
of the derivation, to be checked, not a consequence of it.

A nonce-based cipher~\cite{Rogaway2004} reaches the same place by a different
route: two encryptions under a correct nonce never share a keystream. NPCR is defined for any two ciphertexts of equal length, but for such a
cipher it isolates plaintext sensitivity only if the nonce is held fixed,
which is a misuse confined to the measurement; Section~\ref{sec:xcm} measures a nonce-based
cipher this way and shows what the fixed nonce exposes. Keyed synthetic-IV
constructions derive the nonce from the plaintext under the key and are secure
by design~\cite{rfc5297}; the concern in this paper is not with that practice
but with what NPCR can then measure.


\section{Three regimes of NPCR}
\label{sec:regimes}

NPCR counts the ciphertext bytes that change when one plaintext byte
changes~\cite{Wu2011npcr}. The value that papers report, about $99.61\%$, is the
mean of that count when the two ciphertexts behave like independent uniform
strings. A cipher can reach the same mean for three different reasons, and the
reported number does not say which. This section gives the distribution of NPCR
in each case. The mean turns out to be a poor discriminator; the spread across
trials is a good one.

\subsection{Setting}
In propositions and proofs NPCR is a fraction; in tables and in the text
around them it is given in percent, so that $1/L$ in a proposition appears as
$100/L$ in a table.

Let a ciphertext be a string of $L$ bytes, counted over all colour channels.
Many papers instead report NPCR for each colour channel separately; $L$ is then
the number of pixels in one channel, and the regime C spread is that of a
channel, $\sqrt{3}$ times the whole-image value for three channels. Wherever the
two counts differ we say which one a table uses; in the audit tables
`by channel' marks the per-channel count.
Two plaintexts differ in exactly one byte. We write $j\in\{1,\dots,L\}$ for the
position of that byte in the order in which the diffusion layer scans the data,
after any permutation that depends only on the key, and $\delta$ for the
difference at that byte, taken in the group the diffusion layer uses. NPCR is
the fraction of positions at which the two ciphertexts differ. When a test
chooses the changed pixel uniformly at random and the permutation depends only
on the key, $j$ is uniform on $\{1,\dots,L\}$; we use that model for the
moments below.

A bit flip fixes $\delta$ up to sign. Flipping bit $b$ of a byte changes it by
$\pm 2^{b}$ under addition modulo $256$, and by $2^{b}$ under XOR. For a group
element $\delta\neq 0$ we write $q=\operatorname{ord}(\delta)$ for its additive
order: $q = 256/\gcd(\delta,256)$ under addition, so $q = 2^{8-b}$ for a flip
of bit $b$, and $q=2$ for every nonzero $\delta$ under XOR.

\subsection{Regime A: no propagation}

\begin{definition}[Regime A]
A cipher is in regime A when it is a permutation of positions followed by a
position-wise bijection of byte values, and neither step depends on the
plaintext. Repeating such rounds stays in regime A, because the composition is
again a permutation followed by a position-wise bijection.
\end{definition}

\begin{proposition}[Regime A]
\label{prop:regimeA}
In regime A a one-byte change in the plaintext changes exactly one ciphertext
byte, so NPCR $=1/L$ with zero spread. If each plaintext byte is split across
at most $r$ output bytes by a fixed bit-level rearrangement, NPCR $\le r/L$.
\end{proposition}

\begin{proof}
The permutation moves the changed byte to one position. The position-wise
bijection maps unequal values to unequal values and equal values to equal
values, so exactly that position differs. A fixed rearrangement of bits sends
the changed bits to at most $r$ output bytes.
\end{proof}

XOR with a keystream after a row and column shuffle is a common design, and
so is a modular sum with a fixed mask. Both fall here, whatever the number of
rounds, whenever the keys do not depend on the image;
Sections~\ref{sec:audit:cannot} and~\ref{sec:audit:own} give five published
examples. Regime A is not in itself a defect. A stream cipher such as
ChaCha20~\cite{RFC8439} is in regime A under a fixed key and nonce, since
$C = P \oplus S(K,N)$, and its security rests on the nonce never repeating,
not on plaintext propagation. What regime A does mean is that a differential test under fixed material
measures exactly the absence of propagation between positions, and that a
scheme whose material never changes is a permutation with an independent
position-wise substitution, which falls to plaintext attacks~\cite{LiLo2011}.

\subsection{Regime B: one round of linear chaining}

\begin{definition}[Regime B]
Let $(G,+)$ be one of the groups $(\mathbb{Z}_{256},+)$ and
$(\mathbb{Z}_2^{8},\oplus)$. A diffusion layer is in regime B when it applies one forward and one backward
chaining pass,
\begin{align}
c_i &= p_i + k_i + c_{i-1}, & i&=1,\dots,L,\\
d_i &= c_i + k'_i + d_{i+1}, & i&=L,\dots,1,
\end{align}
with fixed boundary values $c_0$ and $d_{L+1}$, and with masks $k$, $k'$ that
do not depend on the plaintext. A permutation of positions may precede the
layer, and a key-only rearrangement may follow it.
\end{definition}

\begin{proposition}[Exact count in regime B]
\label{prop:regimeB}
Let $m = L-j+1$. In regime B the number of ciphertext bytes that do not change
is
\begin{equation}
Z(j,\delta) \;=\; \left\lfloor \frac{m}{q} \right\rfloor
               \;+\; (j-1)\,\mathbf{1}[\,q \mid m\,],
\label{eq:zcount}
\end{equation}
and NPCR $= 1 - Z(j,\delta)/L$. The count depends on $L$, $j$ and $\delta$
only. It does not depend on the key, the masks or the image.
\end{proposition}

\begin{proof}
Write $\Delta x$ for the difference between the two runs, taken in $G$. The
masks and all plaintext bytes except byte $j$ agree, so the forward pass gives
$\Delta c_i = 0$ for $i<j$ and $\Delta c_i = \Delta c_{i-1} = \delta$ for
$i\ge j$. The backward pass gives $\Delta d_i = \Delta c_i + \Delta d_{i+1}$
with $\Delta d_{L+1}=0$. For $i\ge j$ this is $\Delta d_i = (L-i+1)\,\delta$,
which is zero exactly when $q$ divides $L-i+1$; as $i$ runs from $j$ to $L$,
$L-i+1$ runs over $1,\dots,m$ and $\lfloor m/q\rfloor$ of these values are
multiples of $q$. For $i<j$ the forward difference vanishes and
$\Delta d_i = \Delta d_j = m\,\delta$, so all $j-1$ of those bytes are
unchanged when $q\mid m$ and all change otherwise. A key-only rearrangement
after the layer moves bytes without changing how many differ.
\end{proof}

We call the event $q \mid m$ a \emph{cancellation}: the difference carried into
the head of the image is zero, and the NPCR falls to roughly half its usual
value. The masks drop out of the proof, which is why the result does not depend
on the key. The group matters only through $q$.

\begin{corollary}[Moments in regime B]
\label{cor:moments}
With $j$ uniform and $L\to\infty$, NPCR in regime B has mean $1-1/q$ and
variance $(1-1/q)/(3q)$. A cancellation occurs with probability $1/q$.
Conditional on no cancellation, NPCR is uniform on $[1-1/q,\,1]$, with mean
$1-1/(2q)$ and standard deviation $1/(q\sqrt{12})$. Conditional on a
cancellation, its mean is $(1-1/q)/2$.
\end{corollary}

\begin{proof}
Put $u = m/L$, uniform on $(0,1]$ in the limit. Without cancellation
\eqref{eq:zcount} gives NPCR $\approx 1-u/q$; with cancellation it gives
NPCR $\approx u(1-1/q)$. Mixing the two with weights $1-1/q$ and $1/q$ gives
the stated mean. For the variance, $E[\mathrm{NPCR}^2] = a\bigl(a +
1/(3q^2)\bigr) + a^2/(3q)$ with $a = 1-1/q$, and subtracting $a^2$ leaves
$a/(3q^2)+a^2/(3q) = a/(3q)$.
\end{proof}

Table~\ref{tab:signatures} reports these limits. Exact finite-$L$ moments
follow by averaging \eqref{eq:zcount} over the $L$ positions. With $m$
running over $1,\dots,L$, the tail term contributes $\sum_m \lfloor m/q\rfloor$
and the cancellation term $\sum_{q\mid m}(L-m)$; when $L = kq$ these are
$q\binom{k}{2}+k$ and $q\binom{k}{2}$, so the mean number of unchanged bytes
is $(L-q+1)/q$ and
\begin{equation}
\mathbb{E}[\mathrm{NPCR}] \;=\; 1-\frac{L-q+1}{qL}
 \;=\; \Bigl(1-\frac{1}{q}\Bigr)\Bigl(1+\frac{1}{L}\Bigr),
\label{eq:finiteL}
\end{equation}
which converges to $1-1/q$ from above. At $L=256^2$ and $q=256$ the exact
mean is $99.6109\%$ against the limit $99.6094\%$, a difference of $0.0015$
points; the exact standard deviation is $3.591$ against $3.601$ points, and
the no-cancellation mean $99.8055\%$ against $99.8047\%$.

\subsection{Regime C: independent ciphertexts}

\begin{proposition}[Regime C]
\label{prop:regimeC}
If the two ciphertexts are independent and uniform on bytes, then
$L\cdot\mathrm{NPCR}$ is binomial with parameters $L$ and $255/256$. NPCR has
mean $1-1/256$ and standard deviation
\begin{equation*}
\sigma_C(L)=\sqrt{\frac{1}{256}\cdot\frac{255}{256}\cdot\frac{1}{L}}.
\end{equation*}
\end{proposition}

The binomial law needs more than uniform-looking output. Uniform marginal
bytes give each position a match probability of $1/256$ and hence the mean;
the variance $\sigma_C^2$ and the binomial shape also need the match
indicators to be independent across positions, which a flat histogram does
not establish. This mean and
variance are the ones Wu, Noonan and Agaian~\cite{Wu2011npcr} derived and turned
into a hypothesis test with critical values; we restate them because the other
two regimes are defined against them. Regime C is a probability law for
the pair of ciphertexts, not a construction, and a mechanism does not
establish it. Three kinds of scheme are often expected to come close to it: one that
uses a fresh nonce for each encryption; one whose keys are derived from the
plaintext, for instance from its hash, so that a one-byte change replaces the
whole keystream; and a deterministic cipher with enough rounds of mixing.
Whether a given scheme actually does is a question for the measurements. In the first two
cases the encryption material itself changes between the two runs, and the
measurement then exercises the hash, the nonce and the keystream generator,
and it cannot isolate what the diffusion layer contributes. We therefore keep two properties apart
throughout: whether the material changes in the experiment, which is read from
the protocol or the code, and whether the reported values are compatible with
the independent-uniform reference, which is read from the table. The same
stream cipher is in regime A under fixed material and may be compatible with
regime C under fresh material; the classification belongs to the experiment as
much as to the cipher.

The same binomial law governs a comparison that is sometimes reported in place
of NPCR: the fraction of positions at which a plaintext and its own ciphertext
differ. If the ciphertext is uniform and independent of the plaintext, each
position agrees with probability $1/256$ whatever the plaintext distribution,
so this plain-versus-cipher value also has mean $1-1/256$. It reproduces the
regime~C number for any cipher whose output looks uniform, including one in
regime A.

\subsection{What the mean hides}

\begin{corollary}[The mean does not identify the regime]
\label{cor:blind}
For a flip of the least significant bit under addition ($q=256$), the regime B
mean converges to the regime C mean, $1-1/256 \approx 99.6094\%$, as
$L\to\infty$, exceeding it by $(1-1/q)/L$ at finite $L$ when $q$ divides $L$. The spreads differ by
more than two orders of magnitude over all trials, and by a factor of about
five to twenty ($4.6$ at $256^2$, $18.5$ at $1024^2$) when no cancellation
occurs.
\end{corollary}

The corollary is stated for $q=256$ because that is the case that matters in
practice. A flip of a higher bit under addition, or any flip under XOR, gives
a regime B mean of $99.2\%$ or below, down to $50\%$ for $q=2$, values no
author would report as passing; those rows of Table~\ref{tab:signatures} show how
strongly the result depends on the bit, not a second way to be fooled. The
mean is blind in exactly the configuration that the common test uses: a
least-significant-bit flip on an additive design. One qualification is
needed. A run without cancellation has mean $99.805\%$, which a two-sided test
against $99.609\%$ would also flag; in practice it is not flagged, because a
value above the ideal is read as stronger diffusion rather than as a
departure from independence. The spread makes the same run legible without that misreading, and in the
unconditional case, where the means coincide, the spread, the minimum and the
shape of the distribution are what remain.

Table~\ref{tab:signatures} lists the signatures. Two features matter in
practice. First, which bit is flipped decides the result in regime B: flipping
the top bit gives a mean of $50\%$, and flipping bit~1 gives a regime B
cancellation in one trial in $128$. A paper that does not state the bit does
not define its measurement. Second, with $q=256$ a cancellation is rare
enough that a typical run of $100$ trials contains none; the probability is
$(255/256)^{100}\approx 0.68$. Such a run reports a mean near
$99.805\%$ and a standard deviation near $0.113$ percentage points, which is
the no-cancellation signature, and it looks like a cipher that beats the ideal
value.

\paragraph{Fixed position}
The moments above assume the changed pixel is placed uniformly. Many papers
instead change one fixed pixel, often the first or the centre, in each of
several images. Under a fixed key and a fixed position, $j$ is the same in
every image of the same size, and regime B then returns the \emph{same} NPCR
for every image: a spread of exactly zero across same-size images, which is
a sharper signature than any in Table~\ref{tab:signatures}. Regime C is
unaffected by the choice of position. A fresh key per image changes the
permutation, but it makes the scan position $j$ uniform only if the keyed
permutation family places a fixed input position uniformly; that is an
assumption about the family, and it can fail
(the published permutation of Section~\ref{sec:audit:batch} places a fixed input position measurably closer to the start of the scan than uniform).

\paragraph{How many values are needed}
The audit of Section~\ref{sec:audit} works with published tables of five to
thirty values, often printed to two decimals. We simulated the sample standard
deviation of $n$ regime B values without cancellation against $n$ regime C
values, rounding both to two or to four decimals, and set the threshold at the
$99$th percentile of the regime C spread. With $n=5$ at $256\times256$ the
regime B run exceeds the threshold in $98\%$ of simulations; with $n=5$ at
$512\times512$, or $n\ge10$ at either size, in more than $99.8\%$. Rounding
to two decimals changes these figures by less than one point. Five values therefore suffice under the conditions simulated: independent
trials at random positions, no cancellation, values reported individually
rather than as averages, and a threshold set for one image size. A table
whose entries are themselves means over several positions has a smaller
spread, by about $\sqrt{k}$ for $k$ averaged trials when the trials are
independent with a common variance; a table that mixes image sizes changes
the regime C reference; and a table whose entries share one key still gives
independent observations if the positions are drawn independently, but not
if the position is shared too. The audit records which case applies to each
paper, and where it cannot be determined we say so rather than assign a
regime.

\paragraph{UACI}
The same difference profile fixes UACI. In regime B the backward difference at
position $i\ge j$ is $(L-i+1)\delta$ and at $i<j$ the constant $m\delta$, all
modulo $256$. For a uniform ciphertext byte $c$ and a fixed offset $d$ the
expected absolute change is $\mathbb{E}\lvert((c+d)\bmod 256)-c\rvert =
2d(256-d)/256$, so a tail whose offsets cycle through all residues averages to
the ideal $33.46\%$, while the prefix of length $j-1$ contributes a single
offset $m\delta \bmod 256$ whose expected change ranges from $0$ to $50.2\%$.
Unlike the NPCR count, this is a model, not an identity: it takes the
ciphertext bytes as uniform and independent across positions, and it
attributes the variation across trials to the position $j$ alone. Under that
model, with $\delta=\pm1$, the UACI of a single trial has mean $33.46\%$ and
standard deviation $8.6$ points, and a run of $100$ trials has a mean near
$33.4\%$ with a spread near $8.6$ points; regime C gives $33.46\%$ with a
spread below $0.1$ points at these sizes. The variation from the ciphertext
values themselves is of order $1/\sqrt{L}$ and is negligible beside the
prefix term. We give the model for $\delta=\pm1$ under addition only; other
bits and XOR follow the same argument with their own offset sequences. A
reported UACI spread of several points beside an NPCR near $99.8\%$ is
therefore consistent with regime B and not with regime C.

\begin{table*}[t]
\centering
\caption{NPCR signatures (percent). Regime B values are the $L\to\infty$ limits
of Corollary~\ref{cor:moments}; when $q$ divides $L$ the exact finite-$L$ mean
is $100(1-1/q)(1+1/L)$, $0.0015$ points above the limit at $L=256^2$. ``No
canc.'' conditions on no cancellation. The random-change row averages over a
change $\delta$ uniform on $1,\dots,255$, as in tests that add a random
increment to one pixel; its mean is $100(1-4/255)$.
Regime C values follow from Proposition~\ref{prop:regimeC}. Regime A gives
$100/L$ with zero spread.}
\label{tab:signatures}
\begin{tabular}{llcccc}
\toprule
Regime & Case & Mean & SD & Mean, no canc. & SD, no canc.\\
\midrule
B, addition & bit 0 flipped ($q=256$) & 99.609 & 3.60 & 99.805 & 0.113\\
B, addition & bit 1 flipped ($q=128$) & 99.219 & 5.08 & 99.609 & 0.226\\
B, addition & bit 7 flipped ($q=2$)   & 50.000 & 28.87 & 75.000 & 14.43\\
B, XOR      & any flip ($q=2$)        & 50.000 & 28.87 & 75.000 & 14.43\\
B, addition & random nonzero change & 98.431 & 7.93 & 99.302 & 1.98\\
\midrule
C & $L = 256^2$           & 99.6094 & 0.0244 & -- & --\\
C & $L = 512^2$           & 99.6094 & 0.0122 & -- & --\\
C & $L = 1024^2$          & 99.6094 & 0.0061 & -- & --\\
C & $L = 768\times512\times3$ & 99.6094 & 0.0057 & -- & --\\
\bottomrule
\end{tabular}
\end{table*}

\subsection{Scope}

Proposition~\ref{prop:regimeB} needs chaining that is linear in a single group.
We checked the boundary directly. Forward and backward XOR chaining with fixed
masks matches \eqref{eq:zcount} with $q=2$ exactly in $400$ random trials. The
mixed cell $c_i = k_i\oplus\bigl((p_i+k_i)\bmod 256\bigr)\oplus c_{i-1}$,
the diffusion of Fridrich-type designs~\cite{Fridrich1998,Li2013modxor}, does
not: a modular sum does not carry a fixed XOR difference, and run in both
directions it departs from the formula by up to $99.9$ points. It has an
empirical signature of its own, which we report so that a common
classical cell is covered: over $600$ trials with fixed random masks and a
least-significant-bit flip, NPCR averages $99.64\%$ with a spread of $2.5$
points, and $1.2\%$ of trials fall below $99\%$, so this cell also produces
occasional collapses and a spread far above regime C. Nonlinear feedback, in
which the keystream depends on the previous ciphertext byte, and several
rounds separated by keyed permutations also fall outside the closed form.
Those designs need to be measured, and the signatures of
Table~\ref{tab:signatures} still tell a reader which regime the measured
values resemble.


\section{An audit of published schemes}
\label{sec:audit}

We applied the signatures of Section~\ref{sec:regimes} to twenty-one published
image ciphers. Table~\ref{tab:evidence} records, for each paper, where the
classification was read, where the values were taken from, the perturbation
convention, the image sizes and the randomness policy of the test. The
evidence is of four kinds, and Table~\ref{tab:audit} labels each row with the
kind it rests on: an algebraic classification from
the equations as printed, which is a proven consequence of the specification;
a structural reimplementation of the diffusion layer with random masks in
place of the original generator, which validates the analysed layer but is
not an end-to-end reproduction (for one batch cipher, a structural model of the
whole chain); a comparison of the mean and spread of the
paper's own reported values with the signature of the assigned regime; and,
in one case, an explanation from the printed design that we have not run.
The schemes were chosen in September 2026 from four sources: recent
permutation--diffusion designs that report NPCR (2022--2026), older designs
that later cryptanalysis or evaluation papers discuss (2008--2016), four
image-cipher designs published in this journal (2020--2026), and our own
three. Every scheme examined for the audit is reported here; none was set
aside. The set is selected, not sampled, and its counts say nothing about how
often each regime occurs in the literature. The conclusions these support
differ in strength, and we keep three apart. A
contradiction between the structure as printed and a reported value
(Sections~\ref{sec:audit:cannot} and~\ref{sec:audit:own}) holds whatever the
sampling. A table that is statistically unusual under stated assumptions
(Sections~\ref{sec:audit:c} and~\ref{sec:audit:uaci}) depends on those
assumptions. A table that a model or an implementation of the design rarely
reproduces (Section~\ref{sec:audit:batch}) depends on the model, the
sampling protocol and the implementation as well. Three of the schemes are
our own earlier work.

\begin{figure*}[t]
\centering
\includegraphics[width=\textwidth]{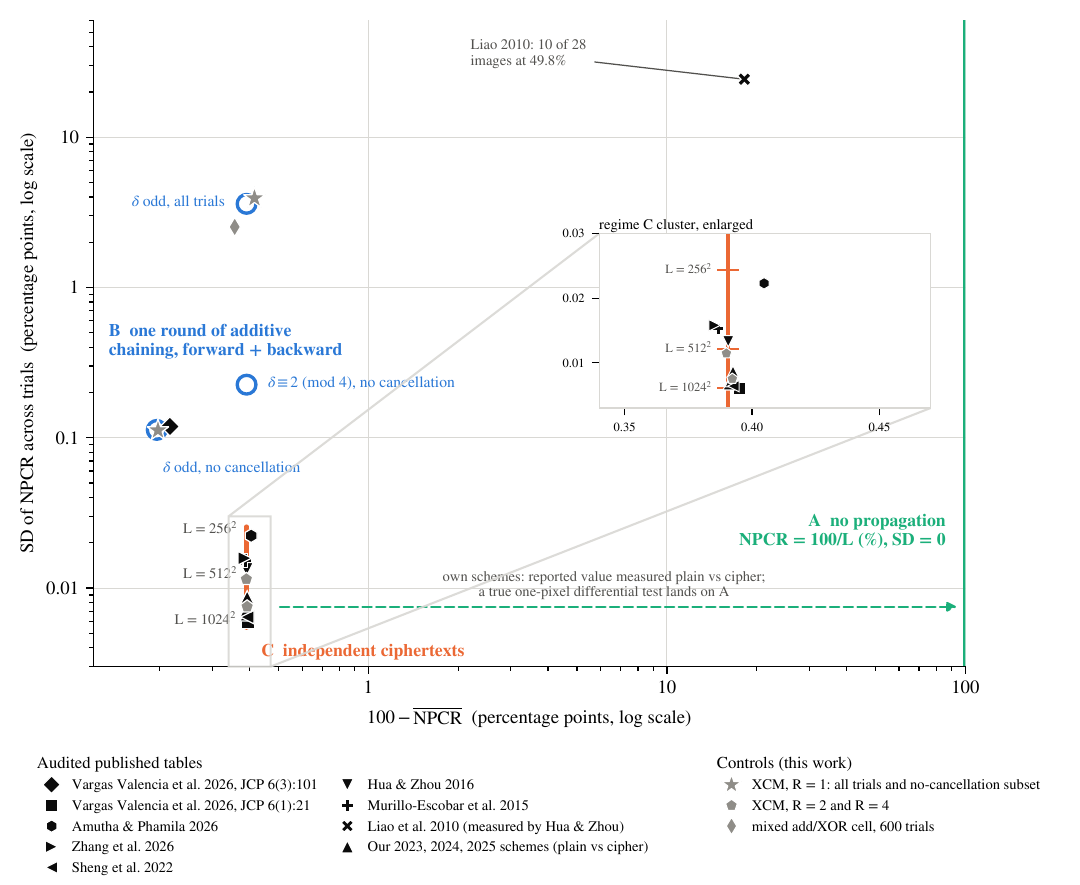}
\caption{Published NPCR reports against the three regime signatures. The
horizontal axis is $100$ minus the mean NPCR, the vertical axis the standard
deviation across the reported values, both in percentage points on log scales.
Regime C is a vertical segment whose height depends on the image size (the
inset enlarges the cluster); regime B, for an odd difference, appears once for
all trials and once for runs with no cancellation; regime A lies off scale at
the right, with zero spread. Most published summaries cluster at the regime C reference; the one regime
B scheme and the bimodal Liao measurements lie outside it. XCM at $R=2$ and $R=4$ and the empirical mixed-cell signature are shown
for comparison; schemes that report a single value (Fu, Wong), only
extremes and averages (Jiang), averages over several trials (Patro, Huang) or
one trial per batch (Song) cannot be placed. Our three earlier schemes report a plain-versus-cipher value that sits
on regime C, while a one-pixel differential test places them on regime A.
Standard deviations computed from values published to four decimal places
are approximate.}
\label{fig:regimes}
\end{figure*}

Table~\ref{tab:audit} summarises the outcome and Fig.~\ref{fig:regimes} places
each report on the plane of NPCR shortfall against spread. The reported values
cluster where regime C predicts, whatever regime the specification puts the
scheme in. Two reports separate from that cluster: the regime B scheme, through
its spread, and Liao's self-adaptive scheme, through two clusters that its
design explains.

\begin{table*}[!t]\centering
\caption{Audit evidence. For each scheme: where the classification was read
(equations, algorithm or section), where the reported NPCR values were taken
from, the paper's stated perturbation, the images and sizes behind the
reported values, and the randomness policy of the differential test.
``n.s.'' means not stated in the paper.}
\label{tab:evidence}\scriptsize\setlength{\tabcolsep}{3pt}
\begin{tabular}{@{}>{\raggedright\arraybackslash}p{2.55cm}>{\raggedright\arraybackslash}p{2.35cm}>{\raggedright\arraybackslash}p{1.7cm}>{\raggedright\arraybackslash}p{2.75cm}>{\raggedright\arraybackslash}p{2.65cm}>{\raggedright\arraybackslash}p{3.15cm}@{}}\toprule
Scheme & Classified from & Values from & Perturbation & Images, size & Material in the test\\ \midrule
Vargas Valencia 2026~\cite{VargasValencia2026jcp101} & Eqs.~14--17, Sec.~4.2, Table~1 & Tables~12, 13 & one bit, all three channels, bit n.s.; 10 random positions per image & 10 USC-SIPI colour, $256^2$ to $576\times787$ & masks from SHA-256 of the key; same key both encryptions\\
Vargas Valencia 2026~\cite{VargasValencia2026jcp21} & Secs.~2.3.1, 2.3.2, App.~A & Table~2 & one pixel, $\Delta=1/255$, random position & one $1024^2$ greyscale, 5 trials & PBKDF2 of key and salt (user strings plus random characters); ``identical key and salt'' in the test\\
Amutha and Phamila 2026~\cite{Amutha2026} & Eqs.~5--7, confusion levels 1--2 & Table~17 (also Table~16) & ``one pixel change'', bit n.s. & 5 medical, $256^2$ (6 USC-SIPI in Table~16) & fixed initial values stated in text\\
Abu-Faraj 2023~\cite{AbuFaraj2023} & Alg.~1, Fig.~4 & Table~4 & none: plain image vs cipher & 12 Kodak, $768\times512$ colour & key only\\
Al-Hyari 2024~\cite{AlHyari2024Access} & Fig.~4, Alg.~1 & Table~9 & none: plain image vs cipher & 12 Kodak & key only\\
Al-Hyari 2025~\cite{AlHyari2025} & Alg.~1, Sec.~3 & Table~3 & none: plain image vs cipher & 12 Kodak & key only\\
Sheng 2022~\cite{Sheng2022} & Secs.~4.1, 4.3, Fig.~8 & Table~5 & one pixel, bit n.s. & Baboon, Peppers, Boat, Sailboat by channel; $256^2$ & SHA-512 of the image\\
Zhang 2026~\cite{Zhang2026sym} & Eqs.~17--21, Sec.~3 & Table~7 & one pixel, bit n.s. & 6 USC-SIPI & SHA-256 of the image\\
Jiang 2026~\cite{Jiang2026jisa} & Secs.~2.1--2.4, Algs.~1--7 & Tables~2, 7, 8 & one channel of one random pixel, random increment & 6 videos, $512^2$ frames, by channel & SHA-256 of each frame; random keys\\
Patro 2020~\cite{Patro2020jisa} & Sec.~3, steps 3, 9--12 & Table~13 & n.s. & 8 greyscale, $512^2$, two batches of four & SHA-256 of the combined batch\\
Lu 2025~\cite{Lu2025jksu} & Alg.~1, Sec.~4.2 & none with spread & n.s. & -- & SHA-256 of the image\\
Anujaa 2025~\cite{Anujaa2025} & Sec.~3 (D1--D6) & none with spread & one bit, position n.s. & -- & SHA-512 of the image\\
Alkhonaini 2024~\cite{Alkhonaini2024} & Steps~2, 6 & none with spread & one bit at (4,7) of Lena & $256^2$ & pixel sum\\
Nkapkop 2014~\cite{Nkapkop2014} & Eq.~10, Alg.~fragments 1--3 & none with spread & n.s. & Lena $512^2$ & pixel sum of shuffled image\\
Fu 2012~\cite{Fu2012} & Eqs.~4--7, Fig.~5 & Table~3 (one value per $R$) & one bit, lower-right corner & Lena $256^2$ & Chebyshev parameter fed by previous cipher pixel\\
Huang 2025~\cite{Huang2025jisa} & Sec.~4, Eq.~13, Algs.~1--2 & Table~6 & one pixel, one channel, change of $1$; $100$ tests & 4 colour, $256^2$ and $512^2$ & fixed keys\\
Song 2024~\cite{Song2024jisa} & Sec.~3, Algs.~2--3 & Table~7 & one stated pixel per batch, least significant bit & 9 USC-SIPI (8 listed) and 4 COCO, mixed sizes & fixed key\\
Hua and Zhou 2016~\cite{HuaZhou2016} & Eqs.~6--7, Sec.~4.2 & Table~2 & one bit, random pixel & 28 USC-SIPI, $256^2$ to $1024^2$ & random surrounding pixels per encryption; redrawn or fixed n.s.\\
Murillo-Escobar 2015~\cite{MurilloEscobar2015} & Eqs.~2--3, 11--12, Table~1 & Table~5 & one bit & $512^2$ colour, 3 values & keystream from pixel sum\\
Wong 2008~\cite{Wong2008} & Eqs.~2--5 & one value & one bit, lower-right corner & $512^2$ & logistic map fed by previous cipher pixel\\
Liao 2010~\cite{Liao2010} & abstract and design description; values from~\cite{HuaZhou2016} & \cite{HuaZhou2016} Table~2 & one bit, random pixel & 28 USC-SIPI & one half of the image encrypts the other\\
\bottomrule\end{tabular}\end{table*}

\subsection{A regime B diffusion layer reproduced exactly}
\label{sec:audit:b}

Vargas Valencia et al.~\cite{VargasValencia2026jcp101} diffuse each colour
channel with one forward and one backward pass of addition modulo $256$
(their Eqs.~14--17), then reorder the three channel values at each pixel with a
key-dependent permutation. The masks come from SHA-256 of the user's key and do
not depend on the image. The scheme is therefore in regime B with $L$ equal to
the number of pixels per channel, and the inter-channel step is a key-only
rearrangement.

We implemented the diffusion and channel-permutation layers from the printed
equations, with masks drawn uniformly at random in place of the paper's
SHA-256-seeded chaotic generator, since Proposition~\ref{prop:regimeB} says
the masks cannot matter, and ran the layer against the proposition for every
bit position: for each of the eight bits, eight random positions in each of
three $512\times512$ images and one $576\times720$ image, sizes taken from
the paper's test set, $32$ trials per bit and $256$ in all. The unchanged-byte count equalled \eqref{eq:zcount} on
every trial. This validates the analysed layer; it is not a reproduction
of the original key schedule.

The paper's differential test changes one pixel of each of ten images at ten
random positions, flipping ``the same bit'' in all three channels, and reports
the mean and standard deviation over all $100$ trials. Because the
flipped bit is the same in every channel, the three channels share one pattern
of unchanged bytes, and the channel permutation only moves identical patterns
around. That explains why the paper's Table~12 reports the same value,
$99.7838\pm0.1190\%$, for the red, green and blue channels.

The paper does not say which bit was flipped. We replicated the protocol
$20{,}000$ times with the closed form for each choice. With the least
significant bit, $67.1\%$ of replications contain no cancellation, and among
those the mean of the $100$ NPCR values averages $99.8049\%$ with a standard
deviation of $0.0113$ percentage points across replications. The reported mean
lies $1.9$ of those standard deviations below, and the reported spread of
$0.1190$ sits at the $89$th percentile of the replicated spreads. Flipping bit
1 would have given a mean near $99.22\%$ and, because a cancellation then
occurs in one trial in $128$, a spread above one point in most runs of $100$;
flipping bit 7 would have given a mean near $50\%$. The published value is what
a least-significant-bit test of this cipher produces. It exceeds the ideal
$99.6094\%$ because of the regime B structure, not because diffusion is
stronger than ideal.

The paper's UACI confirms the reading. Its Table~13 reports
$32.6068\pm7.8677\%$ over the same $100$ trials. Regime C would give
$33.46\%$ with a spread below $0.1$ points. Regime B with a least-significant
bit flip gives, over $4{,}000$ replications of the protocol, a run mean of
$33.44\%$ and a run spread with median $8.6$ points; the reported mean and
spread sit at the $17$th and $16$th percentiles of those replications. A UACI spread of eight points beside an NPCR near $99.8\%$ is what the
constant prefix offset of the regime B difference profile produces, and it is
not what regime C produces; we know of no other mechanism in this design that
would give it, and the paper describes none.

\subsection{Reported values inconsistent with the printed specification}
\label{sec:audit:cannot}

Two schemes are in regime A as printed, yet report NPCR values with the mean
and spread of regime C.

The first~\cite{VargasValencia2026jcp21} applies a key-driven partition and
permutation and then adds a fixed diffusion mask to each pixel, with no
chaining. Keys come from PBKDF2 of a user key and a salt that the paper describes as
``unique and deterministic'' and builds from four user-supplied strings and
an eight-character random sequence, with a further random block appended if
too many characters repeat. That random part would change the material
between encryptions if it were redrawn, but the differential test is stated
to run ``under identical key and salt conditions'', with the same derived key
``ensuring identical chaotic sequences and diffusion masks'', so the two
ciphertexts of each trial share their masks. The diffusion step adds the mask $Y$ to the permuted image $X$
position by position, $Z = X + Y$ reduced to $[0,255]$, with no term from a
neighbouring output. By Proposition~\ref{prop:regimeA}, a one-pixel change in a $1024\times1024$ image
should change one ciphertext pixel, an NPCR of about $0.0001\%$. The paper
reports five trials averaging $99.605\pm0.006\%$. Two independent
$1024\times1024$ ciphertexts would give $99.6094\pm0.0061\%$, and all five
reported values lie within $1.9$ standard deviations of that mean. The cipher itself
is not released, and the released measurement script leaves the encryption
steps to the operator, so we cannot tell which step of the measurement
differs from the specification. What we can say is that the scheme
as printed does not produce the numbers reported for it.

The second~\cite{Amutha2026} uses fixed initial values (for instance
$y_{03} = 0.56543$), two levels of position permutation, an additive mask
reduced modulo $256$ that adds one chaotic matrix entry to each pixel (their
Eq.~6, $S = (I_2 + R) \bmod 256$, again with no neighbouring term), and a
bitwise OR that joins the upper four bits of each pixel in one image quadrant
with the lower four bits of a pixel in another (Eq.~7). One plaintext byte
reaches at most two ciphertext bytes, so NPCR $\le 100\cdot2/65536 \approx 0.0031\%$
on its $256\times256$ test images. The paper reports $99.5953\pm0.0223\%$ over
five medical images (its Table~17) and $99.5999\pm0.0218\%$ over six USC-SIPI
images (its Table~16); regime C at that size predicts $99.6094\pm0.0244\%$.

\subsection{Our own earlier schemes}
\label{sec:audit:own}

Three of our earlier papers~\cite{AbuFaraj2023,AlHyari2024Access,AlHyari2025}
share one structure: a row shuffle and a column shuffle driven by a logistic
map, then XOR with a chaotic key matrix, repeated for four rounds. The keys come
from the user's key alone, and in the 2023 and 2025 schemes one key matrix
per round is applied to the red, green and blue channels alike. The schemes are
in regime A. Four rounds of permutation and
XOR compose to a single permutation followed by a single mask,
$C=\pi(P)\oplus K^{*}$, and our reimplementation of the structure confirms
the consequences. A one-bit change in one pixel of a $768\times512$ colour
image changes one byte of $1{,}179{,}648$, an NPCR of $8.5\times10^{-5}\%$.
Encrypting an all-zero image reveals $K^{*}$. Because the same row and column
shuffle is applied to all three channels, one further image whose three
bytes at each pixel encode that pixel's index, which covers up to $2^{24}$
pixels, reveals $\pi$; with those two chosen plaintexts any ciphertext
decrypts without the key. For a general byte permutation the same attack
needs $1+\lceil\log_{256}L\rceil$ queries, and it is a special case of the
plaintext attacks on permutation-only ciphers~\cite{LiLo2011}. We verified
the two-query version on a structural model of the schemes with the same
shuffle-and-XOR structure; we did not reimplement the original key
generators. Where one key matrix serves all three
channels, the XOR of two ciphertext channels is a rearrangement of the XOR of
the corresponding plaintext channels.

The NPCR values in those papers, $99.609\pm0.007\%$, $99.608\pm0.009\%$ and
$99.608\pm0.007\%$ over twelve Kodak images each, were computed between each
plain image and its own ciphertext. As Section~\ref{sec:regimes} shows, that
comparison yields the regime C value for any cipher whose output looks
uniform, and it does not test diffusion. We report the error here because the
same measurement appears widely in the literature.

\subsection{Changing material, and designs outside the closed form}
\label{sec:audit:c}

Six schemes derive their keys from the plain image: SHA-512 of the
image~\cite{Sheng2022,Anujaa2025}, SHA-256 of the
image~\cite{Zhang2026sym,Lu2025jksu}, or the sum of the pixel
values~\cite{Alkhonaini2024,Nkapkop2014}; a seventh derives its keystream from
the plaintext sum~\cite{MurilloEscobar2015}. In each case a one-pixel change alters the derived material, so the two
ciphertexts are encrypted under different keystreams and the reported NPCR
mostly reflects that replacement, not propagation under fixed material. This
is the first of the two properties in Table~\ref{tab:audit}, and it is read
from the construction. The second, whether the values are compatible with
independent uniform ciphertexts, is not implied by the first and has to be
read from the table. For the four schemes that report no spread there is
nothing to read. Where values are reported, most sit at the C reference:
$99.6147\pm0.0158\%$ over six images for Zhang et al.~\cite{Zhang2026sym},
for instance. Sheng et al.~\cite{Sheng2022} are the exception, in the other
direction. Their eight values have a spread of $0.0064$ points, where eight
independent single trials at $256\times256$ would give about $0.0244$; a
spread this small arises in about $0.05\%$ of such tables printed to four
decimals ($p=0.0005$ from $200{,}000$ simulated tables). That is not evidence against the cipher. Values averaged
over several trials, values that share trials, or a protocol other than the
one described would each produce it, and the paper does not state which
applies. It does make eight independent single trials of the stated test an
unlikely source for the table; what the table does show is a mean at the C
reference. Hua and Zhou~\cite{HuaZhou2016} add
random values around the image at each encryption; if those values are drawn
afresh for the second ciphertext, as the design intends, the material changes
by fresh randomness, and their reported $99.6094\pm0.0133\%$ over
$28$ images of $256^2$ to $1024^2$ pixels is compatible with regime C. The paper
does not say whether the values were held fixed for the test, and if they
were, the value measures two rounds of bit-level permutation and XOR
chaining, which lies outside the closed form.

Patro et al.~\cite{Patro2020jisa} hash the combined batch of images with
SHA-256 and derive their initial values from the digest, so they belong with
the schemes above. Their eight per-image averages, $99.6193\pm0.0047\%$ at
$512\times512$, all lie above the reference, and the paper counts that in the
scheme's favour. We report that pattern descriptively. Under regime C the
chance of a printed average lying above the reference is not exactly one half:
the binomial lattice and rounding to four decimals move it, and by an amount
that depends on how many trials each average holds, which the paper does not
state. For $1$ to $100$ trials it is $0.503$ to $0.508$; for $1{,}000$ and
$10{,}000$ trials it is $0.526$ and $0.581$, because averages then crowd near
the reference and most of them print just above it. As a conditional
sensitivity analysis only, eight of eight above would have probability
$0.004$ to $0.013$ over that range if the averages were independent. They
need not be: the images were encrypted in two batches of four under one
digest per batch, and averages that share a batch can share their
randomness. What holds regardless is that an NPCR above the
reference is not stronger diffusion, and that a systematic excess is what
residual structure produces (Section~\ref{sec:regimes}). With the digest held
fixed the diffusion is a forward-only XOR, first of rows and then of columns,
so a one-pixel change alters exactly the rectangle of ciphertext below and to
the right of its permuted position, a quarter of the image on average over
positions. The reference-level NPCR came from the digest; the linear core
beneath it is what Singh et al.~\cite{Singh2024jisa} later inverted without
the key.

Huang et al.~\cite{Huang2025jisa} add each byte to the previous one and XOR
the sum with a keystream byte, for three forward rounds with a wrap-around,
and then shuffle pixels and bit planes; the closed form does not apply. Their
twelve per-channel values, $99.6111\pm0.0013\%$, are means over $100$ tests on
$256^2$ and $512^2$ colour images and all lie at or just above the reference, by
at most $0.0045$ points. Here the averaging count is stated, so the chance of
a printed value above the reference can be computed: $0.504$ at $256^2$ and
$0.508$ at $512^2$ for means of $100$ trials, and twelve of twelve above would
have probability of about $0.0003$ if the values were independent (the
chance of that one outcome; the two-sided sign test gives $p=0.0005$). They need
not be, since the three channels of one image come from the same encryptions
and how the tests were divided among images is not stated, so the table cannot say more.

Two schemes feed the ciphertext back into the keystream: Fu et
al.~\cite{Fu2012} perturb the Chebyshev parameter with the previous cipher
pixel, and Wong et al.~\cite{Wong2008} drive a logistic map with it. This is
a diffusion mechanism, not a key derivation, and it is nonlinear, so the
closed form does not apply. Fu et al.\ report a single value, $99.61\%$, and
Wong et al.\ $99.36\%$; neither gives a spread to test, although Wong et
al.'s value, at $512\times512$, lies $20$ standard deviations below what
independent ciphertexts give. Fu et al.'s own
table also records a useful baseline: their ``conventional'' one-round
column, a single forward pass with a fixed key, gives $15.74\%$, which is the
single deterministic value that one forward pass with the change near the end
of the scan produces.

\subsection{A structure the spread exposes}
\label{sec:audit:liao}

Hua and Zhou~\cite{HuaZhou2016} also measured the scheme of Liao et
al.~\cite{Liao2010} on $28$ images. Ten values lie between $49.80\%$ and
$49.83\%$ and the rest near $99.61\%$, giving $81.83\pm24.30\%$. None of the
three regimes predicts a tight cluster at $49.8\%$: in regime B the cancelled
trials average $49.8\%$ but spread over the whole interval, and regime C never
produces them. The printed design does predict it. Liao et al.'s scheme is
self-adaptive: the image is split into two halves, and each half is encrypted
using the data of the other. Depending on which half holds the changed pixel
and the order in which the halves are processed, a one-pixel change either
replaces the other half entirely while its own half changes at one pixel,
giving $0.996\times50\%\approx49.8\%$, or propagates through both halves,
giving $99.6\%$. Those are the two clusters. The mean alone, $81.8\%$, would
have read as weak diffusion; the spread shows two behaviours, and the design
explains them. We have not run the scheme, so this reading rests on its
printed description.

\subsection{An NPCR table at the reference beside a UACI table far from it}
\label{sec:audit:uaci}

Jiang et al.~\cite{Jiang2026jisa} remove diffusion on purpose. Each video
frame is permuted and XORed with a keystream whose initial conditions come
from the SHA-256 hash of the frame, so a one-pixel change replaces the
keystream; the authors state that the hash, not diffusion, supplies the
plaintext sensitivity. With the hash held fixed the scheme is in regime A;
with it free, its material changes with the frame, the mechanism of the
schemes in Section~\ref{sec:audit:c}, and its NPCR table is compatible with
regime C. Over six videos
at $512\times512$, three channels each, the eighteen per-channel averages lie
between $99.6084\%$ and $99.6103\%$, and the most extreme single frame is
$4.0$ standard deviations from the reference, which is unremarkable among the
few hundred frames per video that such tables summarise (the paper does not
give the counts).

The UACI table printed beside it is not consistent with regime C. For independent uniform ciphertexts of this
size the UACI of one frame has a standard deviation of $0.046$ points, so over
a few hundred frames the extremes should lie within about $0.15$ points of
$33.46\%$. The reported extremes, printed to six decimals, are $32.5867\%$ and
$34.1402\%$, $19.0$ and $14.6$
standard deviations from the reference, and fifteen of the eighteen rows reach
beyond five; the widest ranges are in the Train and Waterfall videos. The
paper itself quotes Wu's $5\%$ interval for this size, $33.3730$ to
$33.5541\%$, and judges its frames against it; the extremes lie $0.79$ and
$0.59$ points outside that interval. By chance about one frame in twenty is
expected outside it, but such frames should miss it by hundredths of a point,
not by more than half a point. NPCR
records only whether a byte changed, while UACI weighs how far it moved and so
depends on how the ciphertext bytes are distributed. The paper's own histogram
test points the same way: all eighteen per-channel averages of its $\chi^2$
statistic for encrypted frames exceed $255$, the uniform-byte expectation,
ranging from $259.2$ to $287.7$, while an average over, say, $300$
independent frames of uniform bytes would scatter by about $1.3$ (the frame
counts are not given, so this figure is only illustrative). Two kinds of
dependence have to be kept apart here. Dependence between frames or between
channels leaves each frame's own UACI law unchanged; it alters how averages
and extremes over frames are distributed, so it bears on the $\chi^2$
averages and on how many extreme frames to expect, but it cannot move a
single frame nineteen standard deviations. A frame that far out departs from
the per-frame reference itself, which assumes uniform ciphertext bytes that
are independent within the frame and between the two ciphertexts of the pair.
Non-uniform bytes would do that, and so would dependence within the frame or
a measurement that differs from the one described; the tables cannot tell
these apart. We have not run the scheme, and this is a comparison of the published tables
with regime C (Appendix~\ref{app:audit}), not a diagnosis of the cause. It
is the situation Check~1 of Section~\ref{sec:protocol} is written for: the
NPCR mean passed, and the UACI spread printed beside it was the evidence that
the measured ciphertexts do not behave as independent uniform strings.

\subsection{A batch cipher whose reported values are uncommon under its design}
\label{sec:audit:batch}

Song et al.~\cite{Song2024jisa} encrypt a batch of images of different sizes
under one key. After a key-only permutation across the batch, a single chain
runs through every pixel of every image, position by position and image by
image, with the update $S \leftarrow k \oplus ((S + k) \bmod 256) \oplus
S_{\mathrm{prev}}$, first forwards and then backwards (their Algorithm~3 and
Section~3.2).
This is the mixed cell of Section~\ref{sec:regimes} with fixed masks, and it
keeps the property of regime B that decides the table: unless a cancellation
occurs, every position the chain visits before the changed pixel differs,
because the backward pass carries one nonzero difference through all of them.
An image whose positions all precede the change should therefore show
$100\%$. The paper changes
one stated pixel in each of two batches, the least significant bit of a pixel
of Lena and of a COCO image, and reports twelve per-image values between
$99.589\%$ and $99.628\%$, averaging $99.608\%$ and $99.611\%$; none is
$100\%$. Each batch is one trial, so the spread the signatures read is not
available, and the table can only be checked against the structure.

We examined the design in two ways (Appendix~\ref{app:audit}). The first is a
surrogate model: random plaintexts, uniform random permutations and masks in
place of the chaotic ones, the batch composition of the paper's Tables~3
and~4, and the stated pixels. The second is a port of the authors' public C
implementation (\url{https://github.com/TcSong/CrossImgEncryption}, linked
from~\cite{Song2024jisa}). Before using the port we compiled the authors'
three algorithm files unmodified and ran them beside it on small and full-size batches under
the key of the authors' \texttt{main.c} and random keys: the permutation coordinates, both keystreams, the forward pass and
the ciphertext agreed byte for byte in all $17$ runs. We then ran the port on
test images of the stated sizes under $1{,}000$ keys per batch, with initial
values drawn uniformly from $(0.05,0.95)$; the code prescribes no key
distribution, so the frequencies below are conditional on that choice.
Widening the range to $(0.001,0.999)$ changed them little (below). The
paper states the changed pixels by row and column, and the code loads bitmap
rows bottom first, so we counted the row in both directions. For the USC-SIPI
batch, the mean over the listed images was as low as reported in $17$ of
$1{,}000$ surrogate runs and in $2$ and $8$ of $1{,}000$ port runs, one count
for each row convention; for the COCO batch the counts were $12$, $13$ and
$11$. Every listed value was at or below the largest reported one in $3$, $2$
and $8$ USC-SIPI runs and in $4$, $4$ and $1$ COCO runs, in the same order.
With keys drawn from $(0.001,0.999)$ instead, and the row counted in the
displayed image, the mean was as low as reported in $3$ and $13$ runs of
$1{,}000$ and every value at or below the largest in $2$ and $4$. With counts this small the $95\%$ intervals
are wide, from under $0.1\%$ to $2.7\%$ across the twelve frequencies
(Appendix~\ref{app:audit}), but none reaches $3\%$. Low tables arise when the
permutation places the changed pixel near the start of the chain or a
cancellation intervenes. The two batches were encrypted under the same key, so
we do not combine the two frequencies.

The port also shows that the order in which the code visits the images
matters, which the surrogate, ordering them by length, got wrong. With a
least-significant-bit change the difference that enters the backward pass is
odd at the changed cell and at every cell after it, so from the changed cell
onwards a cell with an odd number of cells between it and the end of the
chain, inclusive, cannot cancel. Over most of the COCO batch the chain visits
the four images in turn and the chain length is even, so two of the four
images never cancel after the change; cells before it all differ unless the
difference has cancelled at the changed cell itself. In the code's file-name
order, which we take to be the order the paper lists, the smallest image is
one of the two, and it reached exactly $100\%$ under $996$ and $999$ of
$1{,}000$ keys in the two row conventions; the paper reports $99.611\%$ for it.

The reported values are therefore uncommon under both models of the printed
design, for the key populations tested and our reading of the image order,
and the evidence goes no further than that. That is
a statement about the models, not a demonstration that the authors'
measurement differs from their code: the images are not redistributed, the
authors' compiler and floating-point library may take the chaotic
trajectories elsewhere, and the pixel convention and the order of the COCO
images are our reading of the paper. A reader should treat the table as
unexplained by the design rather than as evidence for it.

\begin{table*}[t]
\centering
\caption{The audit. Two properties are recorded separately. ``Fixed'' is the
regime of the structure as printed with all encryption material held fixed
(``out'': outside the closed form; ``--'': not analysed). ``Material'' says
whether the material changes between the two encryptions of the test, and how;
this is read from the paper's protocol, not from its numbers. ``vs C'' says
whether the reported table is compatible with the independent-uniform
reference of regime C: yes, no (with the reason), or -- where the table cannot
decide. Reported NPCR is mean $\pm$ SD over the values the paper lists, with
their number in parentheses. Predicted is the value the fixed-material regime
gives at the paper's image size. Evidence: A (algebraic classification), S
(structural reimplementation or model), T (comparison with the paper's
table), D (explanation from the printed design, not run). Fig.~\ref{fig:workflow}
and Section~\ref{sec:intro} count each scheme once, under its main finding:
Patro et al.\ and Hua and Zhou with the schemes whose material changes, and
Liao et al.\ with the two-cluster table.}
\label{tab:audit}
\footnotesize\setlength{\tabcolsep}{2.5pt}
\begin{tabular}{@{}>{\raggedright\arraybackslash}p{3.3cm}>{\raggedright\arraybackslash}p{3.25cm}c>{\raggedright\arraybackslash}p{1.55cm}>{\raggedright\arraybackslash}p{2.55cm}>{\raggedright\arraybackslash}p{1.95cm}>{\raggedright\arraybackslash}p{1.75cm}l@{}}
\toprule
Scheme & Structure & Fixed & Material & Reported (\%) & Predicted (\%) & vs C & Ev.\\
\midrule
Vargas Valencia et al.\ 2026~\cite{VargasValencia2026jcp101} & one round, add.\ chaining & B & fixed & $99.7838\pm0.1190$ ($100$) & $99.805\pm0.113$ (bit 0, no canc.) & no: spread & A,S,T\\
Vargas Valencia et al.\ 2026~\cite{VargasValencia2026jcp21} & permutation, unchained sum & A & fixed & $99.605\pm0.006$ ($5$) & $0.0001$ & yes$^{*}$ & A,T\\
Amutha and Phamila 2026~\cite{Amutha2026} & permutation, unchained sum, nibble recombination & A & fixed & $99.5953\pm0.0223$ ($5$) & $\le 0.0031$ & yes$^{*}$ & A,T\\
Abu-Faraj et al.\ 2023~\cite{AbuFaraj2023} & shuffle then XOR, four rounds & A & fixed & $99.609\pm0.007$ ($12$)$^{\dagger}$ & $8.5\times10^{-5}$ & --$^{\dagger}$ & A,S,T\\
Al-Hyari et al.\ 2024~\cite{AlHyari2024Access} & shuffle then XOR, four rounds & A & fixed & $99.608\pm0.009$ ($12$)$^{\dagger}$ & $8.5\times10^{-5}$ & --$^{\dagger}$ & A,S,T\\
Al-Hyari et al.\ 2025~\cite{AlHyari2025} & shuffle then XOR, four rounds & A & fixed & $99.608\pm0.007$ ($12$)$^{\dagger}$ & $8.5\times10^{-5}$ & --$^{\dagger}$ & A,S,T\\
Sheng et al.\ 2022~\cite{Sheng2022} & keys from SHA-512 of image & -- & hash & $99.6073\pm0.0064$ ($8$) & -- & no: spread too small & A,T\\
Zhang et al.\ 2026~\cite{Zhang2026sym} & keys from SHA-256 of image & -- & hash & $99.6147\pm0.0158$ ($6$) & -- & yes & A,T\\
Jiang et al.\ 2026~\cite{Jiang2026jisa} & SHA-256 of each frame, no diffusion & A & hash & $99.6095$ ($18$ averages)$^{\S}$ & $100/L$ & NPCR yes, UACI no & A,T\\
Patro et al.\ 2020~\cite{Patro2020jisa} & SHA-256 of batch; forward XOR of rows, columns & out & hash & $99.6193\pm0.0047$ ($8$ averages) & rectangle, ${\approx}25$ & all above$^{\P}$ & A,T\\
Lu et al.\ 2025~\cite{Lu2025jksu} & keys from SHA-256 of image & -- & hash & -- & -- & -- & A\\
Anujaa et al.\ 2025~\cite{Anujaa2025} & keys from SHA-512 of image & -- & hash & -- & -- & -- & A\\
Alkhonaini et al.\ 2024~\cite{Alkhonaini2024} & keys from pixel sum & -- & pixel sum & -- & -- & -- & A\\
Nkapkop et al.\ 2014~\cite{Nkapkop2014} & keys from pixel sum & -- & pixel sum & -- & -- & -- & A\\
Fu et al.\ 2012~\cite{Fu2012} & ciphertext-fed keystream & out & fixed & $99.61$ ($1$) & -- & yes (one value) & A\\
Hua and Zhou 2016~\cite{HuaZhou2016} & random pixels added per encryption & out & fresh$^{\ddagger}$ & $99.6094\pm0.0133$ ($28$) & -- & yes & A,T\\
Murillo-Escobar et al.\ 2015~\cite{MurilloEscobar2015} & keystream from pixel sum & -- & pixel sum & $99.613\pm0.015$ ($3$) & -- & yes & A,T\\
Wong et al.\ 2008~\cite{Wong2008} & nonlinear feedback & out & fixed & $99.36$ ($1$) & -- & no: $-20$ SD & A\\
Huang et al.\ 2025~\cite{Huang2025jisa} & three forward add/XOR rounds, bit-plane shuffles & out & fixed & $99.6111\pm0.0013$ ($12$ means) & -- & excess$^{\P}$ & A,T\\
Song et al.\ 2024~\cite{Song2024jisa} & batch chain, add/XOR, one round & out & fixed & $99.6092\pm0.0094$ ($12$; 2 trials) & this low in $0.2$--$1.7\%$ of runs & -- (see text) & A,S,T\\
Liao et al.\ 2010~\cite{Liao2010}, measured in~\cite{HuaZhou2016} & self-adaptive, half keys half & -- & by half & $81.83\pm24.30$ ($28$) & $49.8$ or $99.6$ by half & no: two clusters & D,T\\
\bottomrule
\multicolumn{8}{@{}p{17.1cm}@{}}{\scriptsize $^{*}$ Compatible with C, but the printed structure cannot produce it (Section~\ref{sec:audit:cannot}). $^{\dagger}$ Plain image against its own ciphertext, not a differential measurement. $^{\ddagger}$ If redrawn for the second ciphertext, which is not stated; with the values held fixed the structure is outside the closed form. $^{\S}$ UACI extremes $32.5867$ and $34.1402$; independent ciphertexts keep them within about $0.15$ of $33.46$ (Section~\ref{sec:audit:uaci}). $^{\P}$ Every value above the reference. Huang et al.\ state $100$ trials per mean, and a sign test gives $p=0.0005$ if the values are independent; Patro et al.\ do not state the count, so no probability is given (Section~\ref{sec:audit:c}).}
\end{tabular}
\end{table*}


\section{A controlled case}
\label{sec:xcm}

The audit relies on published numbers. To test the signatures where every
parameter is known, we use a cipher of our own, XCM-IC, as a measurement
instrument. Its full specification, the proofs about its map, its complete evaluation,
test vectors and harness are in a technical report deposited with the code
and records~\cite{Alfarajat2026artifact,Alfarajat2026artifactv2}; here we need only its structure.

\subsection{The instrument}
\label{sec:cryptosystem}

XCM-IC is a nonce-based cipher for byte-sampled images; its data flow is
drawn in the deposited report. One HKDF extraction~\cite{Krawczyk2010hkdf} per
message, with the nonce as salt, seeds a chaotic generator, an expanding
circulant map specified in the deposited report. Each of $R$ rounds applies a keyed
Fisher--Yates permutation to all $L$ bytes of the image and then one forward
and one backward pass of addition modulo $256$ with two keystream masks. The
result is authenticated with HMAC-SHA-256 under encrypt-then-MAC. At $R=1$ the
diffusion layer is exactly the regime B layer of Section~\ref{sec:regimes},
with $L$ equal to the number of bytes over all channels.

Because the cipher is nonce-based, two encryptions under a correct nonce never
share a keystream, and NPCR would measure only that. We therefore hold the key
and the nonce fixed for every differential measurement below. This is a misuse
confined to the measurement, and it is what isolates the contribution of the
diffusion layer to the statistic. Two corpora are used: $20{,}000$ nonces on one
Kodak image with $L = 1{,}179{,}648 = 4608 \times 256$, and $46$ images from
Kodak and USC-SIPI under five keys each.

\subsection{One round is regime B, on every trial}

Over $20{,}000$ nonces the flipped change is $\delta = +1$, so $q = 256$, and
Proposition~\ref{prop:regimeB} predicts the NPCR of each trial from the
permuted position $j$ alone. It holds on every trial: the unchanged-byte count equals \eqref{eq:zcount}
on all $20{,}000$, the residual between the recorded NPCR and $1-Z/L$ being
floating-point rounding at $10^{-14}$. The cancellation
event occurs in $78$ trials, a rate of $0.390\%$ with a $95\%$ interval of
$[0.31, 0.49]\%$ (Clopper--Pearson), against the predicted $1/256 = 0.3906\%$. The moments assume that the permutation places the change uniformly. The
residues of the tail length modulo $256$ pass a chi-square test
($\chi^2_{255} = 268.7$, $p = 0.27$), and so does the full position in $256$
bins ($\chi^2_{255} = 246.7$, $p = 0.63$); a Kolmogorov--Smirnov test on the
full position gives $D = 0.010$, $p = 0.03$, which rejects uniformity at the
nominal $5\%$ level. The departure is a small shortfall of positions near the
end of the scan: the empirical distribution function lies $1.0$ point above
the uniform one at $0.70$ of the scan, and the last quarter holds $4{,}816$
trials against $5{,}000$ expected. It does not come from index sampling: the
Fisher--Yates step draws each index by rejection, which is exact for uniform
generator words, so any bias would have to come from the words themselves.
With three tests of the same sample, one rejection at $p=0.03$ is weak
evidence. These tests concern the sampled distribution of the position; the
exact per-trial count needs no assumption about it, and only the moment
predictions do.

\begin{figure*}[!t]\centering
\includegraphics[width=\textwidth]{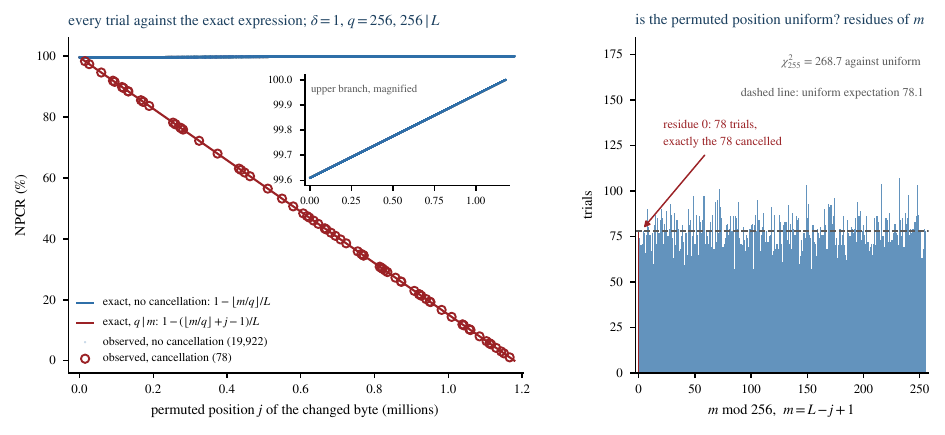}
\caption{Regime B on every trial: $20{,}000$ nonces on one image, $R = 1$,
$\delta = +1$, $L$ divisible by $256$. \emph{Left:} measured NPCR against the
permuted position $j$ of the changed byte, with both branches of
Proposition~\ref{prop:regimeB} overlaid; every point lies on its branch to
$10^{-14}$ and nothing is fitted. The inset shows that the uncancelled branch
rises from $99.61\%$ toward $100\%$. \emph{Right:} trials per residue of
the tail length $m = L - j + 1$ modulo $256$. Residue $0$ is the cancellation
event, $q \mid m$, and holds $78$ trials against a uniform expectation of
$78.1$.}
\label{fig:npcr}
\end{figure*}

The two means behave as Corollary~\ref{cor:moments} predicts. Excluding the
cancelled trials, the mean is $99.803\%$ with a standard deviation of $0.112$
points, against $99.805\%$ and $0.113$. Over all trials the mean is $99.585\%$
against $99.609\%$ and the standard deviation $3.93$ against $3.60$. The
shortfall in the mean comes almost entirely from where the $78$ cancellations
happened to fall: their mean is $43.9\%$ against an expected $49.8\%$, a
deviation of $1.8$ standard errors, which accounts for $0.023$ of the
$0.025$-point shortfall. A sample of this size is dominated by a handful of
cancelled trials, which is why a mean over a few images says little.
Figure~\ref{fig:npcr} shows every trial against the two branches of the
formula.

The cancelled trials show what the appearance metrics miss. The most extreme
of the $20{,}000$ changed $1.10\%$ of the ciphertext bytes. In a parameter
sweep reported in the deposited report, one cancelled cell changed
$10.5\%$ of the ciphertext; among the $96$ cells of that sweep with the same
length it had the lowest maximum adjacent-pixel correlation and the
second-highest entropy, and it passed chi-square with $\chi^2 = 217.9$. Every
appearance measure rated that ciphertext among the best in the sweep.

\subsection{Mask values cancel from the one-round difference}
\label{sec:xcm:generator}

\begin{figure*}[!t]\centering
\includegraphics[width=\textwidth]{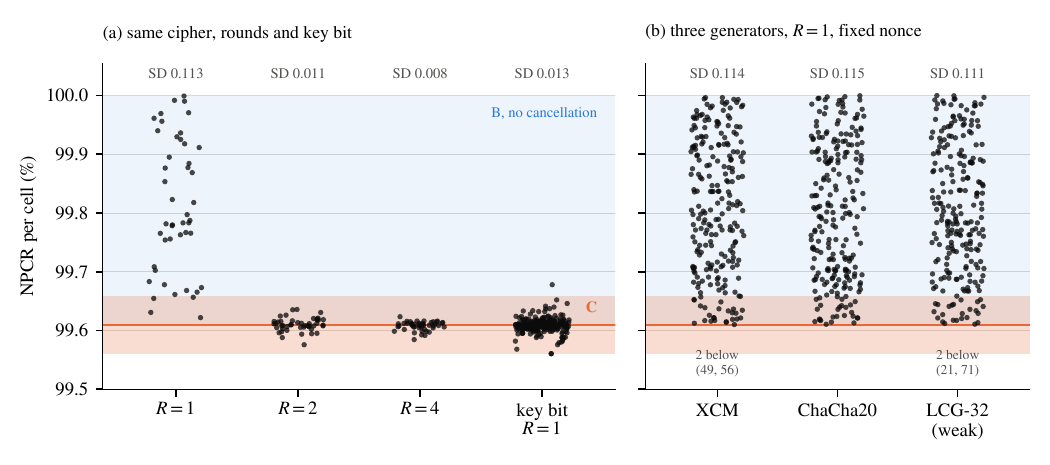}
\caption{NPCR per cell under a fixed nonce. The blue band is the regime B
no-cancellation range $[99.609, 100]\%$; the orange band is regime C, mean
$\pm 2$ standard deviations at the smallest image in the corpus ($256^2$
bytes). \emph{(a)} The same cipher at $R = 1$, $2$ and $4$ over $46$ images,
and with a flipped key bit at $R=1$ over $230$ cells. One round fills the
regime B band; at two rounds, or with a new keystream, this
least-significant-bit test does not distinguish the cipher from regime C (Fig.~\ref{fig:bitsrounds} shows
the other bits). \emph{(b)} Three generators in an identical scaffold at
$R = 1$, $230$ cells each. The LCG, recoverable from one output word, is not
separated from ChaCha20. Cancelled cells lie below the axis and are listed
with their values.}
\label{fig:xcmregimes}
\end{figure*}

Proposition~\ref{prop:regimeB} says the masks drop out of the count, so for a
given scan position the NPCR cannot depend on which generator produced them.
A generator can still act through the permutation, and so through the
distribution of the scan position; on our instrument the chi-square tests of
that position did not reject uniformity while a Kolmogorov--Smirnov test did at
the $5\%$ level (Section~\ref{sec:xcm}), and the comparison below runs the
whole cipher, so it includes that route. We tested
this by replacing the chaotic generator with ChaCha20~\cite{RFC8439} and then
with a $32$-bit linear congruential generator,
$x\mapsto 1664525x+1013904223 \bmod 2^{32}$, whose state is recoverable from a
single output word within a keystream block (the scaffold reseeds every
$512$ words, so prediction across blocks is a separate matter), keeping
everything else identical. Each arm ran $230$ cells
at $R=1$ under a fixed nonce. Excluding cancellations, the three arms give
$99.810\pm0.114\%$ for XCM, $99.802\pm0.115\%$ for ChaCha20 and
$99.805\pm0.111\%$ for the LCG. XCM and the LCG each had two cancellations and
ChaCha20 none, against $0.9$ expected per arm. The three arms carry the regime B signature and are not separated by these
measurements (Fig.~\ref{fig:xcmregimes}b). NPCR measured this way cannot see a generator
that is broken by one output word, because the statistic is a property of the
diffusion layer and not of the keystream.

\subsection{More rounds: C-like statistics for one bit, affine structure for all}
\label{sec:xcm:rounds}

At $R = 2$ the change entering the second round is no longer confined to one
byte, the second permutation scatters it, and the count of
Proposition~\ref{prop:regimeB} no longer applies. The measurements over the
$46$ images, one key each (the generator comparison of
Section~\ref{sec:xcm:generator} uses five), show what replaces it (Fig.~\ref{fig:xcmregimes}a).
At $R=1$ the NPCR is $99.812\pm0.113\%$, the regime B no-cancellation
signature, with no cancellation among the $46$. At $R=2$ it is
$99.610\pm0.011\%$ and at $R=4$ $99.608\pm0.008\%$. Dividing each deviation
from $99.6094\%$ by the regime C standard deviation at that cell's length gives
standardised values with a standard deviation of $1.02$ at $R = 2$ and
$0.82$ at $R = 4$, against $1$ for independent ciphertexts. A flipped key bit
at $R = 1$ replaces the keystream and gives $99.609\pm0.013\%$ over $230$
cells, standardised standard deviation $1.03$. The mean moves by $0.20$
points between one and two rounds; the spread moves by a factor of ten.

Those statistics resemble regime C. The cipher at $R=2$ is not in regime C,
and the difference is visible under a different perturbation. Under a fixed
key and nonce every keystream term cancels in the difference, and the
difference is then a linear function of the plaintext difference over
$\mathbb{Z}_{256}$ at any round count; the encryption itself is affine.

\begin{proposition}[Multi-round difference]
\label{prop:multiround}
Let $U$ be the lower-triangular prefix-sum matrix and $V$ the upper-triangular
reverse-prefix-sum matrix over $\mathbb{Z}_{256}$, and let $P_r$ be the
permutation matrix of round $r$. Under a fixed key and nonce,
\[
  \Delta C_R \;=\; A_R\,\Delta P, \qquad
  A_R = \prod_{r=R}^{1} V U P_r \pmod{256}.
\]
\end{proposition}

\begin{proof}
One round maps a difference $\Delta$ to $VUP_r\Delta$: the permutation
relocates it, the forward pass applies $U$, the backward pass applies $V$, and
every keystream term cancels because the key and nonce are the same in both
encryptions. Composing over rounds gives the product.
\end{proof}

The observation is not new as cryptanalysis: Zhou and Yu~\cite{ZhouYu2026jisa}
use the fact that the ciphertext difference of a two-round additive-chaining
cipher does not depend on its diffusion keys to recover its permutations. Our
use of it is different: it says what a differential table can and cannot show.
Each factor is invertible, so $A_R$ is an invertible linear map over
$\mathbb{Z}_{256}$ at a fixed key, nonce and length; the deposited report checks the law against the
reference implementation for $R\le4$ and sets out what it gives an adversary
under a repeated nonce. Two consequences matter here. First, for a change of
$2^b$ in one byte, every entry of $\Delta C = 2^b A_R e_j$ is a multiple of
$2^b$, at every $R$ and for every permutation sequence. Second, the identity
$E(P+u+v)-E(P+u)-E(P+v)+E(P)\equiv0 \pmod{256}$ holds on the ciphertext body
for any $u$, $v$, since $E$ is affine. A top-bit change can
therefore only ever produce differences in $\{0,128\}$: the low seven bits of
every ciphertext byte are untouched however many rounds are run. This
restricts the values of the differences, not their number; with the change
at the last scan position and $R=1$, for instance, every byte differs by
$128$ and NPCR is $100\%$.

\begin{figure*}[!t]\centering
\includegraphics[width=\textwidth]{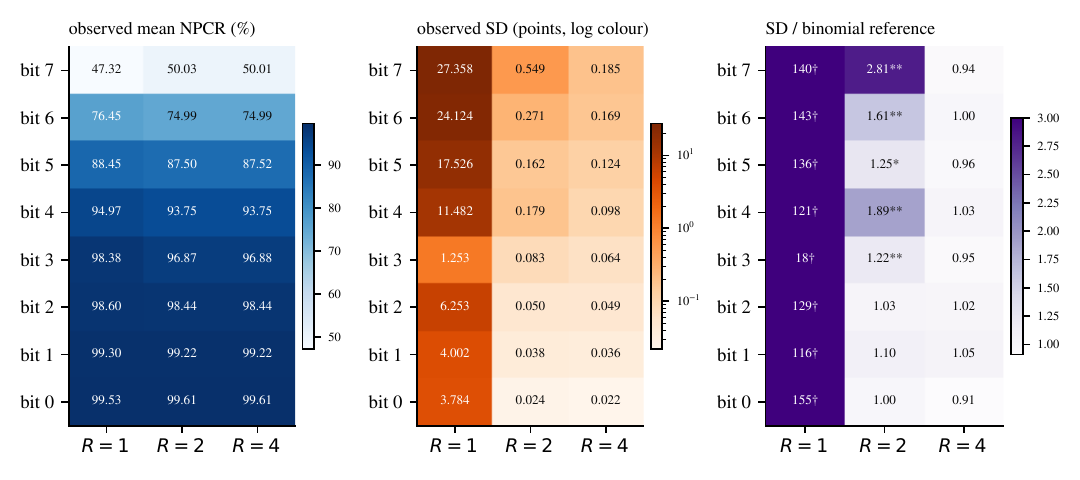}
\caption{NPCR by flipped input bit and round count through the XCM-IC
implementation, $256\times256$ images, $200$ trials per cell with a fresh
key, nonce, image and position each. \emph{Left:} observed mean NPCR.
\emph{Centre:} observed standard deviation across trials, logarithmic colour
scale. \emph{Right:} that standard deviation divided by the
independent-binomial reference $100\sqrt{p_b(1-p_b)/L}$ with $p_b=1-2^{b-8}$.
$\dagger$: $R=1$, where the exact law of Proposition~\ref{prop:regimeB} gives
a ratio of about $148$ and no interval is used. For $R\ge2$, from a BCa bootstrap
over trials: $^{*}$ the pointwise $95\%$ interval for the ratio excludes one;
$^{**}$ so does the interval adjusted for the $16$ cells at $R\ge2$. From
$R=2$ the mean is at $1-2^{b-8}$ for every bit; excess dispersion is detected
at $R=2$ for bits~3 to~7 and not at $R=4$.}
\label{fig:bitsrounds}
\end{figure*}

Figure~\ref{fig:bitsrounds} shows what the implementation does for all eight
bits: $200$ trials per cell at $256\times256$, each with a fresh random key,
nonce, image and position, the key and nonce held fixed within the trial. No
trial violated the divisibility restriction, and the affine identity held on
the ciphertext body at $R=1$, $2$ and $4$ in $30$ further trials. From $R=2$ on the observed mean
sits at $1-2^{b-8}$ for a change in bit $b$, within two standard errors in
every cell and within one in $14$ of the $16$: $99.61\%$ for bit~0, $99.22\%$ for bit~1, $96.87\%$ for bit~3 and
$50.0\%$ for bit~7. The dispersion behaves differently. At $R=1$ it follows the exact law of
Proposition~\ref{prop:regimeB} with the change position uniform, which puts
the standard deviation at about $148$ times the independent-binomial reference
$100\sqrt{p_b(1-p_b)/L}$, $p_b = 1-2^{b-8}$, for every bit at this size. Every one of the
$4{,}800$ trials at $R=1$, over the three image sizes, had the difference
profile of that law, and its unchanged-byte count equalled \eqref{eq:zcount}
at the scan position read from the profile. The observed ratios, $18$ to
$155$ at $256\times256$, scatter widely because a sample of $200$ holds only a
few cancellations, each far below the other values. The counts follow
$200/q$: $585$ cancellations over the $24$ cells against $597.7$ expected. The
lowest ratio, $18$ for bit~3, belongs to the cell with the fewest, one against
$6.25$ expected ($P=0.013$), the smallest of the $24$ cell probabilities; a
value this small is not unusual among $24$ cells. At $R=2$ the spread has fallen by two orders of
magnitude but still exceeds the reference for bits~3 to~7, by factors of
$1.2$ to $2.8$. At $R=4$ no ratio's interval excludes one. The experimental unit is
the trial: each of the $200$ trials in a cell has its own key, nonce, image
and position, so the trials are independent. The NPCR values of a cell are
not normal at $R=2$ (excess kurtosis up to $59$), so a chi-square interval
for the standard deviation, which assumes normality, is only an
approximation there. We therefore resampled the trials (BCa bootstrap,
$20{,}000$ resamples). The pointwise $95\%$ intervals exclude one for bits~3
to~7 at $R=2$ and for no cell at $R=4$; adjusted for the $16$ cells at
$R\ge2$ (Bonferroni), they still exclude one for bits~3, 4, 6 and~7. Bit~1 at
$R=2$, whose ratio is $1.10$, is not among them: its chi-square interval just
excludes one, but its bootstrap interval, $0.99$ to $1.29$, does not.
Among the round counts we tested, the mean looked reference-like at a
smaller count than the dispersion, and the mean stays bit-dependent at every
round count tested. At $R=4$ no excess dispersion was detected at this sample size, which
is not the same as showing that the distributions agree. A
least-significant-bit test at $R\ge2$ therefore returns summaries close to the
regime C mean and spread while the cipher retains a structure the test cannot
see, which is one more reason to read a single
NPCR summary as descriptive rather than as evidence of independence. The
same experiment at $250\times261$, a length not divisible by $256$, and at
$512\times512$ gives the same picture: pointwise excess at $R=2$ for bits~5
to~7 and for bits~1 and~3 to~7 respectively, and no excess at $R=4$, where
a single cell ($512^2$, bit~0) has a pointwise interval below one that the
adjustment for $16$ cells removes (deposited records, which give every
interval).

\subsection{What regime B exposes}

A fixed nonce also shows what the linear structure gives an adversary who can
obtain one. Under a repeated nonce the ciphertext difference of two messages
that differ in one byte is, at $R=1$ and with positions counted from one,
$\Delta C_i = \delta\,(L-\max(i,j)+1) \bmod 256$: a constant prefix followed by
a descending ramp, so the permuted position $j$ of the change is read off
where the prefix ends. With chosen plaintexts, $L-1$ pairs recover the
permutation, and one known plaintext then gives an equivalent additive mask
that decrypts further messages under that nonce; a keyed substitution before
diffusion changes $\delta$ but not the shape. The deposited report
verifies these statements against the reference implementation and extends
them to $R>1$. These attacks require nonce reuse and do not establish an
attack in the nonce-respecting setting. They describe the setting in which
NPCR is measured, and the same algebra applies to every fixed-key regime B
scheme, for which there is no nonce to respect.


\section{Beyond NPCR: what the other measurements miss}
\label{sec:beyond}

NPCR is not the only measurement in the template that can pass without
testing what it is taken to test. The same instrument lets us check two
others: the appearance metrics, against a generator that is known to be
broken, and the chaotic map itself, against its own finite-state structure.

\subsection{Appearance metrics cannot see a broken generator}
\label{sec:eval:weak}

The three-arm scaffold of Section~\ref{sec:xcm} also records the appearance
and image-quality metrics that papers report beside NPCR. Table~\ref{tab:weak}
gives them over $230$ cells per arm, $46$ images under five keys each. Mean
entropy is $7.99932$ in all three arms, identical to five decimal places;
maximum adjacent-pixel correlation differs by at most $9\times10^{-5}$; MSE
differs by $0.03\%$ and SSIM~\cite{Wang2004ssim} by $10^{-4}$. The LCG passes
chi-square in $221$ cells against $217$ for XCM and $220$ for ChaCha20. On these
measures the generator that one output word breaks scores marginally better
than the other two.

\begin{table}[!t]\centering
\caption{Appearance and image-quality metrics for three generators in an
identical scaffold, $230$ cells per arm. The LCG is predictable from one output
word; no metric separates it from the other two generators. Each column is
averaged over cells separately, so the PSNR is the mean of the per-cell
PSNRs, not the PSNR of the mean MSE.}
\label{tab:weak}\small
\setlength{\tabcolsep}{3.2pt}
\begin{tabular}{@{}lrrrrrr@{}}\toprule
Generator & Entropy & $\max|\rho|$ & $\chi^2$ pass & MSE & PSNR & SSIM\\ \midrule
XCM      & 7.99932 & 0.00232 & 217/230 & 9374 & 8.549 & 0.0089\\
ChaCha20 & 7.99932 & 0.00241 & 220/230 & 9377 & 8.548 & 0.0088\\
LCG-32   & 7.99932 & 0.00241 & 221/230 & 9375 & 8.548 & 0.0089\\
\bottomrule\end{tabular}\end{table}

Two features of the design bear on Check~2 of Section~\ref{sec:protocol}.
Each image is encrypted under five independently derived keys, not one. On the
$24$ Kodak images, which share one size, a one-way random-effects analysis of
ciphertext entropy with image as the factor gives $F(23,96)=1.14$, $p=0.32$,
with the image component at $2.7\%$ of the variance and key-to-key variation
at $97.3\%$; a table with one key per image, the usual design, measures key
variation and attributes it to images. And the entropy deficit of a uniform
byte stream is, to leading order, a function of its
length~\cite{Wu2013localentropy}, so entropy figures from images of different
sizes compare the sizes; a weak-generator arm must therefore run on the same
images at the same sizes as the scheme it controls.

\subsection{The map that is analysed is not the map that runs}
\label{sec:eval:graph}

Chaos-based ciphers justify their generators with continuous quantities: a
positive Lyapunov exponent, a bifurcation diagram. The implemented map acts on
a finite set of $w$-bit words, where every orbit is eventually periodic, and
the questions that matter are finite-state ones: what fraction of states is
reachable, how many lie on cycles, how long the tails are. The map inside
XCM-IC updates $n$ words of $w$ bits as
$Y_i = 3X_i + X_{i-1} + \varepsilon\,Q(X_{i-1}) \bmod 2^w$, with
$Q(X_{i-1})$ the truncated fixed-point value of $\bigl(u(1-u)\bigr)^2$ at
$u=X_{i-1}/2^w$ and indices modulo $n$; the cipher uses $n=4$, $w=32$ and $\varepsilon=2$.
Appendix~\ref{app:finite} states the results used here with their proofs and
the enumerated values, following the observables of our finite-state audit of
chaotic maps~\cite{AlfarajatAlhaj2026finite}. Two bear on practice. The linear
core $X\mapsto AX \bmod 2^w$, with $A=3I+S$ and $S$ the cyclic shift, is a
circulant whose Smith normal form over $\mathbb{Z}$ is
$\operatorname{diag}(1,\dots,1,\lvert\det A\rvert)$; its image fraction is
therefore $2^{-\min(w,\,v_2(\det A))}$, which is $1/8$ for $n=2$ and $1/16$
for $n=4$ at every width, and because $n$ is a power of two every state
reaches the zero state within $nw$ steps. (For a general integer matrix the
exponent is a sum over all its invariant factors; the single nontrivial factor
is what makes it this simple here.) The continuous map is expanding, so no
continuous diagnostic can show this collapse. And for even $\varepsilon$ a
parity argument confines the full map to at most half of its state space; the
enumerated image fractions at $\varepsilon=2$, $0.440$ to $0.460$ in every
enumerated cell but the smallest, sit near the value a random map into that half would
give, $(1-e^{-2})/2 = 0.432$. The reference value $\varepsilon=2$ had been
chosen, before this bound was found, by a rule fixed in advance that scored
candidates by closeness to the unrestricted random-mapping value
$1-e^{-1}\approx0.632$, which the bound makes unreachable. An enumeration at
reduced width, which took seconds, would have shown that the criterion was
the wrong one before the rule was written.

\subsection{Generator-level evidence}
\label{sec:eval:batteries}
\label{sec:eval:recovery}

The generator was also run through PractRand, TestU01 and NIST
SP~800-22~\cite{practrand,LEcuyerSimard2007,NIST80022} and a Z3
state-recovery ladder~\cite{deMouraBjorner2008z3}, beside ChaCha20 through the
same derivation. No argument in this paper uses those results, and passing
them would not be evidence of security; the tables are in the deposited
report.


\section{A three-check protocol}
\label{sec:protocol}

None of the three checks needs more than an evaluation already runs: the
first reads trials the paper already has, the second repeats them once with
another generator, and the third took seconds to minutes on our map
(Table~\ref{tab:protocol}). They are diagnostic: they say what a reported table measures, and they can show that a
scheme is not what its numbers suggest. They cannot show that it is secure.

\paragraph{Check 1: report NPCR so that the regime can be read}
Compute NPCR and UACI between two ciphertexts, never between a plaintext and
its ciphertext. State which bit or value was changed, whether positions were
chosen at random or fixed, the number of trials, and the mean, standard
deviation and minimum of each statistic. In the simulated setting of Section~\ref{sec:regimes}, independent
individually reported trials at one image size, five values at two decimals
gave high detection power; other reporting protocols need their own
calibration. State
whether keys, nonces or any other parameter depend on the plaintext. Then
compare the mean and spread with Table~\ref{tab:signatures}. A spread near the
binomial value of regime C, with keys independent of the image, is consistent
with a diffusion layer that mixes for that perturbation; a spread near
$0.113$ points, a UACI spread of several points, identical values across
same-size images under a fixed position, or any value near $50\%$ when a
high bit is flipped, is consistent with regime B and not with regime C; any
cell far below $99\%$ departs from the reference, a cancellation being one
possible cause, and should be reported, not dropped. Where the entries are
averages, share both key and position, or mix image sizes without saying
which, the comparison needs its own calibration and should be reported as
inconclusive rather than forced into a regime; a shared key with
independently drawn positions still gives independent observations
(Section~\ref{sec:regimes}). Report more than one bit: a low-bit result
that matches regime C says nothing about a top-bit result
(Section~\ref{sec:xcm:rounds}). A scheme
whose keys depend on the image should also report NPCR with that dependence
switched off, since otherwise the value mostly reflects the hash. The test of Wu,
Noonan and Agaian~\cite{Wu2011npcr} remains useful for a single value, but its
NPCR part rejects only values that are too low. For a least-significant-bit
flip under one-round additive chaining, with the change position uniform, a
value falls below the reference only when a cancellation occurs, so the
one-sided test passes about $255$ trials in $256$ at any usual threshold.
Applied in both directions it would not;
Section~\ref{sec:protocol:bench} measures what each version finds.

\paragraph{Check 2: run a known-weak generator through the same evaluation}
Replace the scheme's keystream generator with one whose weakness is not in
doubt, such as a $32$-bit linear congruential generator, and repeat every
measurement the paper reports. Any metric that does not separate the two arms
is not evidence about the generator, whatever value it takes. In our scaffold
no appearance metric, and no NPCR measured under a fixed key, separated them
(Sections~\ref{sec:xcm} and~\ref{sec:eval:weak}). Run the control on the same
images at the same sizes, and use several keys per image, so that neither
size nor key-to-key variation is read as an image effect
(Section~\ref{sec:eval:weak}).

\paragraph{Check 3: enumerate the map at a reduced word length}
Before relying on a continuous diagnostic, run the implemented map at a word
length small enough to enumerate, $2^{16}$ to $2^{24}$ states, and report its
image fraction, cyclic-node count and tail lengths against the random-mapping
reference~\cite{FlajoletOdlyzko1990,Lanford1998}. The fraction of states on
cycles is what Guih\'eneuf calls the degree of recurrence of a
discretisation~\cite{Guiheneuf2019recurrence}; we set out these observables
for several maps used in image ciphers in earlier
work~\cite{AlfarajatAlhaj2026finite}. On our map this exposed a degenerate linear core and a parity
bound in seconds (Section~\ref{sec:eval:graph} and
Appendix~\ref{app:finite}), neither of which any Lyapunov exponent or
bifurcation diagram could show.

\begin{table}[!t]\centering
\caption{The three checks: what each costs and what it detects. Check~1 can
be run on a published table with \texttt{npcr\_check}
(Section~\ref{sec:protocol:tool}).}
\label{tab:protocol}\small
\begin{tabular}{@{}>{\raggedright\arraybackslash}p{1.75cm}>{\raggedright\arraybackslash}p{2.75cm}>{\raggedright\arraybackslash}p{3.0cm}@{}}\toprule
Check & Cost & Detects\\ \midrule
NPCR regime & Existing trials; report bit, count, SD, minimum, UACI; one rerun with plaintext dependence off & Regime A or B behind a $99.6\%$ mean; replaced material; plain-vs-cipher values; bit dependence; UACI departing from the reference while NPCR does not\\
Weak generator & One extra arm & Metrics that cannot see generator quality\\
Enumeration & Seconds to minutes at $2^{16}$--$2^{24}$ states & Degenerate cores, parity and other algebraic confinement, short cycles\\
\bottomrule\end{tabular}\end{table}

\subsection{Check 1 as a tool}
\label{sec:protocol:tool}

Check~1 needs no cipher code, so we release it as a short program,
\texttt{npcr\_check}, that an author, reviewer or evaluator can run on a
table before trusting it. Its input is what a paper prints, the values or their
mean, spread and count, the image sizes and whether each value is one trial or
an average, together with two facts read from the equations: the regime, and
whether keys or nonces depend on the plaintext. Its rules are listed in Appendix~\ref{app:checker}. It answers four
questions in order: is the table a differential measurement at all; does the printed
structure allow the reported values; is the encryption material held fixed;
and do the mean, spread, minimum and UACI extremes agree with independent
ciphertexts, with one-round chaining, or with neither. The spread is tested in
both directions, too large and too small, at $0.5\%$ per tail. Rounding matters here:
values printed to two decimals at $1024\times1024$ often coincide, so a zero
spread is common under the null itself, and a coarse rounding step also
fattens the upper tail. Both tails are therefore calibrated on simulated null
tables rounded to the precision the paper prints, and the lower tail is
withheld when the precision is not known. The mean is tested by a $z$-test that also allows for
rounding, and a systematic excess by a sign test, each at the $1\%$ level. The
sign test uses the exact chance that a printed value lies above the
reference, which depends on the lattice, the rounding and the number of
trials behind each value; when that number is not stated it gives no
probability. It also treats the values as independent, which a table does not guarantee
when values share images, batches or encryptions, so its result is reported as
a flag rather than a verdict. UACI extremes are screened at five standard
deviations of the independent-ciphertext law; because the number of frames or
trials behind them is often unstated, this is a screening rule, not a
calibrated test. When a test rejects and nothing in the table points to a
specific structure, the verdict is ``departure from regime C, cause
unresolved''. The checker returns a verdict and the
tests behind it, or says why the table cannot decide: averages, whose spread
is not that of single trials, one trial per batch, a single value, or no table
at all. When none of its tests rejects, the verdict reads ``no discrepancy with
regime C'', which means that the implemented tests found nothing, not that the
values are certified independent. Verdicts of two kinds are kept apart. One
rests on the printed equations (``inconsistent with the printed
specification''). The other names a structure from the table alone
(``compatible with a one-round chaining signature'', ``two clusters''); such a
label is a statistical suggestion, marked as one in the output, and only the
equations can confirm it.

We validated the checker on the simulated papers of
Section~\ref{sec:protocol:bench}, $1{,}000$ per cipher, entered as a paper
would print them: five NPCR values, the UACI extremes, the image size, the
precision, and nothing about the structure. Its tests run separately, so we
measured the combined false alarm directly, on simulated null tables of five
independent single trials at $256^2$, $512^2$ and $1024^2$, printed to four
and to two decimals, each NPCR and UACI pair drawn from the exact joint law of
one pair of independent uniform ciphertexts. It flagged between $1.0$ and
$2.3\%$ in all six
conditions, with the lower-tail test contributing $0.3$ to $0.5$ points where
the precision allows it and none at two decimals on $1024^2$, where rounding
hides the spread. At two decimals on $512^2$ the sign test adds $0.3$ points:
there a single trial prints above the reference with probability $0.64$, so
five values all at or below it are rare enough to reject (counts, intervals and the generation of every table are in
Appendix~\ref{app:checker}, Table~\ref{tab:checkval}). On the $1{,}000$ null papers of the benchmark it flagged $1.6\%$.
It flagged all $1{,}000$ papers of the permutation-and-mask cipher and of
one-round additive, XOR, mixed and S-box chaining ($95\%$ Clopper--Pearson
interval for each rate $99.6$ to $100\%$), and its labels pointed to the right
structure in most: ``limited propagation'' for all of the first, a one-round
chaining signature for $94$ to $97\%$ of the additive, mixed and S-box papers,
and the XOR label for $54\%$ of the XOR papers, the rest being reported as a
departure with the cause unresolved. Table~\ref{tab:confusion} in
Appendix~\ref{app:checker} gives the full matrix of labels against the
simulated mechanisms, wrong suggestions included. Two-round additive
chaining was mostly not flagged at the least significant bit ($4.0\%$ of
papers, against $1.6\%$ for the null), as expected, and was flagged in every
paper when the top-bit table was entered; the hashed variants passed as
regime C and carried the material-change flag that their stated key policy
triggers.

The benchmark ciphers are the cases the labels were written for, so we also
fed the checker $5{,}000$ tables from each of five alternatives outside them
(Appendix~\ref{app:checker}),
at $256\times256$ and four decimals. Averages of ten trials reported as single
trials were flagged in $28\%$ of tables, always as a spread too small; values
sharing a common component (correlation $0.9$) in $45\%$, as too small a spread
or a mean off the reference; independent but non-uniform ciphertext bytes, with
collision probability $1.2/256$, in all of them, almost all as a departure with the cause
unresolved.
The chaining label is the one that could mislead, and it stayed rare where it
does not belong: $0.5\%$ of the non-uniform tables carried it, against $89\%$ of
tables that mix regime B and regime C values and $18\%$ of tables of regime B
values averaged over ten positions, both of which do contain chaining. The two-cluster label needs at least
three values on each side of $99\%$, so five-value tables cannot receive it;
on tables of $10$ and $28$ values it was given to $74\%$ and $99.9\%$ of
tables drawn from two clusters, and to no table of the four other kinds
(null, regime B under addition or XOR, and the B/C mixture; $5{,}000$ tables
each at either size; Table~\ref{tab:clusters}). The
labels therefore narrow the search; they do not replace reading the design.

Run on the twenty-one audited schemes, seventeen of which print a table,
from their printed numbers and stated conditions, its verdicts agree with
Section~\ref{sec:audit} in every row; where the table alone cannot decide, it
gives one of the four reasons above (Table~\ref{tab:verdicts}). It also flags what a reader skimming
the tables would miss: Sheng et al.'s eight values are unusually tightly
grouped for independent single trials (lower-tail $p=0.0005$); all twelve of
Huang et al.'s means of $100$ trials lie above the reference (two-sided sign
test $p=0.0005$, with the chance of a printed value above the reference
computed for that averaging count and rounding, $0.504$ to $0.508$, and if
the values were independent, which the channel structure of the table does
not guarantee); all eight of Patro et al.'s averages lie above it too, which
that paper counts in its favour, but it does not state how many trials each
average holds, so the checker gives no probability
(Section~\ref{sec:audit:c}); and the single value of Wong et al.,
$99.36\%$ at $512\times512$, lies $20$ standard deviations below it.

\begin{table*}[!t]\centering
\caption{Output of \texttt{npcr\_check} on the twenty-one audited schemes, from their
printed numbers and stated conditions (input \texttt{audit\_tables.json}).
SD values are percentage points. For a single image size the regime C SD
is given with a two-sided $p$ (upper or lower tail when one of them rejects;
the lower tail comes from simulated null tables rounded as printed);
for mixed sizes the range of regime C SDs over the sizes is given, and the
tails are tested against its ends; when a tail rejects, the reference at that
end is shown. Where none of the $200{,}000$ simulated null tables reaches the
observed spread, no $p$-value is given: the entry is a $95\%$ upper
confidence bound, $3/200{,}000=1.5\times10^{-5}$, on the tail probability.}
\label{tab:verdicts}\scriptsize
\begin{tabular}{@{}>{\raggedright\arraybackslash}p{4.3cm}>{\raggedright\arraybackslash}p{5.1cm}>{\raggedright\arraybackslash}p{6.6cm}@{}}\toprule
Scheme & Verdict & Tests behind the verdict\\ \midrule
Vargas Valencia et al. 2026 (JCP 6(3):101) & compatible with a one-round chaining signature & SD 0.1190 vs 0.0244, upper: none of 200{,}000 null tables as large (tail probability ${\le}1.5\times10^{-5}$ with 95\% confidence); mean $z=71.6$\\
Vargas Valencia et al. 2026 (JCP 6(1):21) & inconsistent with the printed specification & SD 0.0059 vs 0.0061, two-sided $p=0.89$; mean $z=-1.6$; 1/5 above\\
Amutha and Phamila 2026 & inconsistent with the printed specification & SD 0.0223 vs 0.0244, two-sided $p=1.00$; mean $z=-1.3$\\
Abu-Faraj et al. 2023 & not differential & --\\
Al-Hyari et al. 2024 & not differential & --\\
Al-Hyari et al. 2025 & not differential & --\\
Sheng et al. 2022 & spread too small for independent single trials; material change & SD 0.0064 vs 0.0244, lower $p=0.0005$; mean $z=-0.2$\\
Zhang et al. 2026 & no discrepancy with regime C; material change & SD 0.0158, C range 0.0122--0.0244 (mixed sizes); mean $z=0.5$\\
Jiang et al. 2026 (JISA) & NPCR compatible with C, UACI beyond the 5-SD screen; material change & UACI extreme $z=19$\\
Patro et al. 2020 (JISA) & inconclusive (averages); material change & 8/8 above (no $p$: averaging count unknown)\\
Lu et al. 2025 & no table to check; material change & --\\
Anujaa et al. 2025 & no table to check; material change & --\\
Alkhonaini et al. 2024 & no table to check; material change & --\\
Nkapkop et al. 2014 & no table to check; material change & --\\
Fu et al. 2012 & single value: inconclusive & mean $z=0.0$\\
Hua and Zhou 2016 & no discrepancy with regime C; material change & SD 0.0133, C range 0.0061--0.0244 (mixed sizes); mean $z=0.0$; 16/28 above\\
Murillo-Escobar et al. 2015 & no discrepancy with regime C; material change & SD 0.0153 vs 0.0122, two-sided $p=0.65$; mean $z=0.6$; 2/3 above\\
Wong et al. 2008 & single value outside the regime C range & mean $z=-19.9$\\
Huang et al. 2025 (JISA) & inconclusive (averages); systematic excess & mean $z=2.4$; 12/12 above, sign test $p=0.0005$\\
Song et al. 2024 (JISA) & inconclusive from the table; see structural model & SD 0.0094, C range 0.0061--0.0244 (mixed sizes); mean $z=0.0$; 7/12 above\\
Liao et al. 2010 (in Hua and Zhou) & two clusters: neither regime B nor C; material change & SD 24.3022 vs 0.0244, upper: none of 200{,}000 null tables as large (tail probability ${\le}1.5\times10^{-5}$ with 95\% confidence); mean $z=-3861.1$; 14/28 above; min far below 99\%\\
\bottomrule\end{tabular}\end{table*}

\subsection{The first check against the Wu--Noonan--Agaian test}
\label{sec:protocol:bench}

The first check reads the same table as the test of Wu, Noonan and
Agaian~\cite{Wu2011npcr}, so it is fair to ask what it finds that the test
does not. The procedure compared here, which we call the signature check, is
a calibrated four-statistic rule built for this benchmark. It is not the
released \texttt{npcr\_check}: the checker has more rules, labels as well as
detects, and flags $1.0$ to $2.3\%$ of null tables (Section~\ref{sec:protocol:tool}).
Only the Checker column of Table~\ref{tab:bench} describes the checker; every
other rate below belongs to the signature check or to the Wu tests. We compared the two on nine simulated ciphers whose structure is
known, and on independent ciphertexts as the null (Table~\ref{tab:bench};
the ciphers and the calibration are defined in Appendix~\ref{app:bench}). Each simulated paper uses one key and five
trials, each on a different $256\times256$ test image at a random position,
and prints NPCR and UACI to four decimals, as most papers do; there are
$1{,}000$ papers per cipher. The Wu--Noonan--Agaian test is applied as its
authors define it, to each value, with the per-value level set so that a
paper of independent ciphertexts would be flagged $1\%$ of the time if the
normal approximation were exact; the measured rates are in the first row. The signature
check reads four numbers from the same five values: the minimum NPCR, the
distance of the mean from $99.6094\%$ in either direction, and the standard
deviations of NPCR and UACI, with thresholds set for a $1\%$ family false
alarm on $400{,}000$ simulated null tables. The two-bit columns add five
trials with the top bit flipped, again at a nominal $1\%$ overall, and the last column adds a
rerun with the image-derived key switched off.

The NPCR half of the Wu test, which is the half most papers apply, catches
only the ciphers whose mean is visibly wrong: the permutation-and-mask cipher
and XOR chaining. It flags one-round additive, mixed and nonlinear chaining
in $2$ to $4\%$ of papers, close to its false-alarm rate, because it rejects
only values that are too low. Wu's UACI interval, applied per value, does
catch linear and mixed chaining, through their UACI spread. It misses a
one-round chain with an S-box, whose UACI stays near the reference while its
NPCR values scatter far above and below it; the signature check flags every
one-round structure in every paper from the same five values. Where a column
reads $100.0$, all $1{,}000$ papers were flagged, and the $95\%$
Clopper--Pearson interval for the rate is $99.6$ to $100\%$. Two rounds of
additive chaining are weakly detected at the least significant bit ($1.6$ to
$2.2\%$, against null rates of $0.4$ to $1.0\%$) and
obvious to both at the top bit, where the mean is near $50\%$: for that cipher
what matters is which bits are reported, not which statistic is applied.
For the two hashed-key ciphers of the benchmark, none of the statistics we
tested separated them from independent ciphertexts; switching the dependence
off did, in every paper. With ten
trials per paper the pattern is the same, and the NPCR half of the Wu test
still flags at most $7\%$ of papers for additive, mixed and S-box chaining.
It is also the same with values printed to two decimals, at $512\times512$
and on colour images (Table~\ref{tab:benchrobust} in Appendix~\ref{app:bench}); the one change is that
Wu's UACI interval catches the S-box chain more often as the image grows, in
$34\%$ of papers at $512^2$, against $100\%$ for the signature check in every
condition.

Table~\ref{tab:ablation} separates the ingredients. It applies each one
alone to the same papers, and adds a two-sided version of Wu's NPCR test. As
specified, the tests are nominal $1\%$ tests whose achieved null rates differ
($0.91$ to $1.33\%$ on the exact joint null); to compare them at
approximately matched error rates we also recalibrated every threshold on one
sample of $400{,}000$ null tables and measured the achieved rate on a second,
independent sample ($0.88$ to $1.01\%$; Appendix~\ref{app:bench} gives the
procedure). The detection rates in the table use these recalibrated
thresholds. What the usual test lacks is the
upper side. Wu's own per-value NPCR test, made two-sided, flags every linear
chain in all $1{,}000$ papers and the mixed and S-box chains in $99.9\%$; the
distance of the mean from the reference on its own flags all of them. The
spread alone flags one-round additive, mixed and S-box chaining in $97.6$ to
$98.5\%$ of papers. The full four-rule signature does not improve on the
mean-only rule in this benchmark, and the two-sided per-value test comes within
$0.1$ points of both. Two rounds of additive chaining stay weakly detected:
$2.6\%$ for the mean-only rule against its null rate of $0.99\%$, and $4.0\%$
for the checker against $1.6\%$ (Table~\ref{tab:bench}).
Detection, then, needs only one change to current practice: test in both
directions, so that a significant upper-tail departure of NPCR is read as a
departure, as Section~\ref{sec:regimes} argued, not as stronger diffusion. What the spread adds is the reading of a flagged table.
A mean too high with a spread near $0.1$ points is compatible with a
one-round chaining signature; a spread too small makes independent single
trials unlikely; two clusters point to the design. Those readings are what
Sections~\ref{sec:audit} and~\ref{sec:protocol:tool} rest on, and no
single-value test supplies them. The protocol changes, a second bit and a
rerun with the dependence off, are what catch the two-round and hashed
ciphers, and no statistic of the least-significant-bit table replaces them.

The two-round S-box cipher is flagged by nothing. For a check that reads only
the differential table that is the expected result, and it means no more than
that the table is compatible with regime C; it is not evidence that the
cipher is secure. The simulated ciphers stand in for published designs and the
thresholds assume one image size; a table that mixes sizes needs its own
calibration, as Check~1 says.

\begin{table*}[!t]\centering
\caption{Share of simulated papers flagged (\%), for the Wu--Noonan--Agaian
per-value test~\cite{Wu2011npcr} and the signature check of Check~1.
$256\times256$ test images, one key and five trials per paper, values to four
decimals, $1{,}000$ papers per row; every column except Checker is a nominal $1\%$
family test on independent ciphertexts, with the measured rate in the first
row. Wu N: NPCR test only. Wu N+U:
NPCR and UACI tests. Signature: minimum, mean, and NPCR and UACI standard
deviations. Checker: the released \texttt{npcr\_check} on the same five values,
not tuned to a $1\%$ family rate, share of papers not returned as ``no discrepancy with regime C''
(Section~\ref{sec:protocol:tool}). Two-bit columns add a top-bit flip at each position. Off: the
two-bit signature check plus a rerun with plaintext-derived material switched
off. The null row comes from $1{,}000$ papers; Table~\ref{tab:ablation} gives the null rates on $400{,}000$ tables.}
\label{tab:bench}\small
\begin{tabular}{@{}lrrrrrrr@{}}\toprule
 & \multicolumn{4}{c}{Least significant bit} & \multicolumn{2}{c}{Two bits} & \\
\cmidrule(lr){2-5}\cmidrule(lr){6-7}
Simulated cipher & Wu N & Wu N+U & Signature & Checker & Wu N+U & Signature & Off\\ \midrule
Independent uniform ciphertexts (null) & 0.8 & 1.0 & 0.4 & 1.6 & 1.0 & 0.5 & 0.5\\
\midrule
Permutation and additive mask & 100.0 & 100.0 & 100.0 & 100.0 & 100.0 & 100.0 & 100.0\\
One round, addition (regime B, $q=256$) & 2.4 & 100.0 & 100.0 & 100.0 & 100.0 & 100.0 & 100.0\\
One round, XOR (regime B, $q=2$) & 100.0 & 100.0 & 100.0 & 100.0 & 100.0 & 100.0 & 100.0\\
One round, addition and XOR & 3.5 & 100.0 & 100.0 & 100.0 & 100.0 & 100.0 & 100.0\\
One round, S-box in the chain & 2.6 & 6.4 & 100.0 & 100.0 & 7.9 & 100.0 & 100.0\\
Two rounds, addition & 2.2 & 1.6 & 2.1 & 4.0 & 100.0 & 100.0 & 100.0\\
Two rounds, S-box in the chain & 1.1 & 1.3 & 0.8 & 1.4 & 1.0 & 0.4 & 0.4\\
As A, key from SHA-256 of the image & 1.4 & 1.1 & 1.1 & 2.0 & 1.1 & 0.8 & 100.0\\
As B-add, key from SHA-256 of the image & 1.5 & 1.0 & 0.4 & 2.0 & 1.3 & 0.3 & 100.0\\
\bottomrule\end{tabular}\end{table*}

\begin{table*}[!t]\centering
\caption{Which ingredient detects what: share of the simulated papers of
Table~\ref{tab:bench} flagged (\%), least significant bit, five values per
paper, $1{,}000$ papers per cipher. Null tables come from the exact joint law
of NPCR and UACI for independent uniform ciphertexts. Nominal: achieved
null rate of the test as specified. Recal.: achieved null rate on a
validation sample after recalibration on a separate sample; the detection
columns use the recalibrated thresholds, at approximately matched rates
(procedure in Appendix~\ref{app:bench}). A value of $100.0$ is
$1{,}000$ of $1{,}000$ (Clopper--Pearson $95\%$ interval $99.6$--$100$).}
\label{tab:ablation}\small
\begin{tabular}{@{}lrrrrrrrr@{}}\toprule
 & \multicolumn{2}{c}{Null rate} & \multicolumn{6}{c}{Detected, recalibrated thresholds}\\
\cmidrule(lr){2-3}\cmidrule(lr){4-9}
Test & Nominal & Recal. & Mask & B-add & B-xor & Mixed & S-box R1 & Add R2\\ \midrule
Wu NPCR, one-sided & 1.33 & 0.93 & 100.0 & 2.4 & 100.0 & 3.1 & 2.6 & 1.8\\
Wu NPCR, two-sided & 0.98 & 0.88 & 100.0 & 100.0 & 100.0 & 99.9 & 99.9 & 1.8\\
Wu UACI & 1.00 & 1.01 & 100.0 & 100.0 & 100.0 & 100.0 & 10.6 & 0.6\\
Wu NPCR + UACI & 1.12 & 0.97 & 100.0 & 100.0 & 100.0 & 100.0 & 6.9 & 1.4\\
Wu NPCR two-sided + UACI & 1.02 & 0.94 & 100.0 & 100.0 & 100.0 & 100.0 & 99.9 & 1.6\\
Mean only & 0.96 & 0.99 & 100.0 & 100.0 & 100.0 & 100.0 & 100.0 & 2.6\\
NPCR spread, upper & 1.02 & 1.01 & 0.0 & 97.6 & 100.0 & 98.5 & 98.2 & 1.4\\
NPCR spread, two-sided & 0.99 & 1.00 & 100.0 & 96.8 & 100.0 & 97.9 & 97.8 & 1.6\\
Mean + NPCR spread & 0.97 & 0.99 & 100.0 & 100.0 & 100.0 & 99.9 & 100.0 & 2.3\\
Signature, four rules & 0.91 & 1.00 & 100.0 & 100.0 & 100.0 & 100.0 & 100.0 & 2.1\\
\bottomrule\end{tabular}\end{table*}


\section{Discussion and limitations}
\label{sec:discussion}

\paragraph{Why published values cluster at $99.61\%$}
Figure~\ref{fig:regimes} shows almost every reported value near the regime C
point, whatever regime the specification puts the scheme in; the regime B
scheme separates from the cluster through its spread, and Liao's through its
two clusters. A reader cannot tell
the routes to that point apart from the mean, which is why the spread, the
flipped bit and the material held fixed have to be reported with it. The
reverse inference is weaker: a spread compatible with a regime, under the
stated assumptions, is weak evidence for it and no proof of it, since a nonlinear design, a mixture of protocols or
a shared key and position can move the spread as well. The claims in this
paper are of the form ``the printed structure predicts this, and the reported
table shows it''; where the structure is unknown, the signatures narrow the
possibilities and no more. The
comparison across the audit is coarser than the one we ran on our own data,
since published spreads are taken over images of different sizes and,
usually, different keys; the five-value result of Section~\ref{sec:regimes} holds for
individually reported trials at one size and is not a licence to read every
short table.

\paragraph{What this means in practice}
For an editor or reviewer, a table at $99.61\%$ does not identify the
regime, so on its own it is not the evidence of diffusion that papers present
it as; the items Check~1 asks for cost an author minutes, and with the checker
they settle many cases and say why the rest cannot be settled. For anyone evaluating an image cipher before use, the
regime has direct consequences for an attacker. A scheme in regime A with
key-only material falls to a few chosen images, as our structural model of our
own three shows. A fixed-key scheme with additive chaining, over one round or
more, obeys the linear difference law of Section~\ref{sec:xcm:rounds}, from which Zhou and Yu~\cite{ZhouYu2026jisa}
recovered the permutations of a two-round cipher. A scheme whose
reference-level NPCR comes only from deriving its keys from the image is as
strong as that derivation: a pixel sum can be held fixed by choosing images with equal sums,
which is how Wong et al.~\cite{Wong2024jisa} broke one such scheme, and with a
cryptographic hash the question moves to how the digest is protected and
conveyed to the receiver. For designers, a reference-level NPCR is cheap, as
Jiang et al.~\cite{Jiang2026jisa} show by obtaining it deliberately without
diffusion. There it came from a change of material, and that is a fact about
the protocol and the code, which the NPCR value itself does not establish.

\paragraph{On auditing our own work}
Three of the audited schemes are ours, published under the author's former
name. We found their regime A structure while looking for a regime B example
to test, and we report it because the same measurement error appears widely
and because the chosen-plaintext break follows directly from the structure.

\paragraph{Limitations}
Every statement about an audited scheme is taken from its published paper.
The audit covers twenty-one schemes, selected rather than sampled: recent
permutation--diffusion designs that report NPCR (2022--2026), older designs
that later cryptanalysis or evaluation papers discuss (2008--2016), our own
three, and four image-cipher designs published in the Journal of Information
Security and Applications between 2020 and 2026, one of which a later
cryptanalysis in the same journal has broken. It is not a random sample, and
we make no claim about how often each regime occurs in the literature. The closed form covers one round of chaining that is linear in
one group; when $q$ divides $L$ its finite-length mean exceeds the limit by
$(1-1/q)/L$, which is $0.0015$ points at $256^2$, and its standard deviation falls short of the
limit by $0.01$ points there. Mixed addition and XOR, nonlinear feedback and several rounds with
keyed permutations fall outside it and have to be measured; the signatures
then say which regime the measurements resemble, not why. The UACI results
rest on a uniform-byte model and are stated for one perturbation. That a
scheme's material changes with the image is recorded from its construction;
it is not a proof that its ciphertexts are independent, and a pixel-sum key, for instance, is unchanged by a
compensating two-pixel change. The round-count study of
Section~\ref{sec:xcm:rounds} runs the implementation; its divisibility
statement is algebraic and holds for every permutation sequence, while its
means and dispersions are sample estimates from $200$ trials per cell. Where a reported
value disagrees with its printed specification we cannot say whether the
specification, the implementation or the measurement is at fault, and we have
not tried to. For XCM-IC, confidentiality is not proved, the finite-state
results are exhaustive only up to $2^{24}$ states against $2^{128}$ in the
reference instance, and performance is not competitive: AES-256-GCM on the
same host is $165$ times faster (deposited report). The Liao reading
in Section~\ref{sec:audit:liao} is from the printed design and has not been
run, and the reimplementations in the audit cover the analysed layers, not
the original key schedules. The generator batteries of
Section~\ref{sec:eval:batteries} test the derived, extracted and reseeded
stream, not an uninterrupted orbit of the map.

\section{Conclusion}
\label{sec:conclusion}

NPCR is a descriptive statistic whose interpretation depends on the
encryption structure and on what the comparison holds fixed. For one forward
and backward round of linear chaining an exact count explains both the high
values and the cancellation events, and its asymptotic mean can match the
independent-ciphertext reference while the distribution differs from it.
Additional rounds can produce summaries in which no departure from the
reference is detected for a particular perturbation while algebraic structure
remains: a top-bit input change leaves the lower seven output bits unchanged
at every round count, and among the round counts we tested ($1$, $2$ and
$4$), the mean looked reference-like at a smaller count than the
dispersion did.

Applied to twenty-one published schemes, the count and its signatures
reproduced one diffusion layer and accounted for its reported NPCR and UACI,
found two specifications inconsistent with the values reported for them,
identified three of our own schemes as permutation-plus-mask ciphers whose
reported values were not differential measurements, showed that ten more
change their encryption material with the plaintext or per encryption, so
that their reported NPCR, where given, cannot isolate what diffusion
contributes, found in one of them reported statistics inconsistent with
independent uniform ciphertexts under the stated measurement, found a batch cipher whose printed table is
uncommon under two models of its design, and explained a bimodal table from the design
of a self-adaptive scheme. On a controlled cipher the count
held on $20{,}000$ trials, and neither NPCR nor any appearance metric
separated a generator predictable from one word from ChaCha20.

We ask authors and reviewers for three things: NPCR and UACI between two
ciphertexts, reported with the flipped bit, the material held fixed, the
number of trials and the spread, for more than one bit; one arm with a
known-weak generator on the same images; and an enumeration of the map at a
small word length. None needs more than the evaluation already runs, and
each would have caught at least one of the problems reported here. On
simulated ciphers, testing NPCR in both directions from five values flagged
$99.9$ to $100\%$ of the papers for each one-round chaining structure tested,
including a nonlinear one that the usual one-sided Wu--Noonan--Agaian test
flagged in $2.6\%$; the
spread and the other signatures then helped say which structure produced the
table. The least a reviewer can do is treat a significant upper-tail
departure of NPCR as a finding rather than a merit. The three checks
support reproducible diagnosis within their stated assumptions; they do not
establish ciphertext independence, resistance to differential cryptanalysis,
or confidentiality.

\section*{Data availability}
\label{sec:reproducibility}
The XCM-IC reference implementation, test vectors, the Tier~A and battery
records with their preregistrations and freeze digests are deposited on
Zenodo at \url{https://doi.org/10.5281/zenodo.22796548}~\cite{Alfarajat2026artifact}.
The audit and round-count material of this paper (the reimplementation of the
regime B diffusion layer, the closed-form, power, UACI, mixed-cell and XOR
checks, the eight-bit experiment with its seeds and trial-level records, the
position-uniformity tests, the detection benchmark of Table~\ref{tab:bench}
with its simulated ciphers and calibration, the comparison of
Section~\ref{sec:audit:uaci}, the structural model of
Section~\ref{sec:audit:batch}, the checker \texttt{npcr\_check} of
Section~\ref{sec:protocol:tool} with its input tables and verdicts, the screening record with every reported value in
Table~\ref{tab:audit}, and the figure scripts) is deposited as version~2 of
the same record, at \url{https://doi.org/10.5281/zenodo.23069747}~\cite{Alfarajat2026artifactv2}, with a script that regenerates every table and figure, a
check that recomputes the numbers printed here from the records, and a
technical report with the full XCM-IC specification, the proofs of the
finite-state results and the complete evaluation of the instrument.

\section*{Declaration of competing interest}
The author declares that he has no known competing financial interests or
personal relationships that could have appeared to influence the work reported
in this paper.

\section*{Funding}
This research did not receive any specific grant from funding agencies in the
public, commercial, or not-for-profit sectors.

\appendix

\section{Audit procedure}
\label{app:audit}

\paragraph{Screening}
Each scheme was read in full text. It was assigned to regime A when its
printed equations consist of key-only permutations and position-wise
bijections; to regime B when its diffusion is one forward and one backward
pass of chaining linear in a single group, with image-independent masks.
Designs with nonlinear ciphertext feedback into the keystream, or with
chaining that mixes addition and XOR, were marked as outside the closed form.
Separately, each scheme was recorded as changing its material when a key,
initial value or parameter depends on the plain image or on randomness drawn
per encryption, and as holding it fixed otherwise; this second property was
read from the protocol and the equations, never from the reported numbers.
Compatibility of the reported values with regime C was then judged from the
table alone. Reported values were transcribed from
the papers' own tables, with table numbers, into the deposited screening
record.

\paragraph{Structural reimplementations}
The diffusion layer of Vargas Valencia et al.~\cite{VargasValencia2026jcp101}
was implemented from its Eqs.~14--17 and its inter-channel permutation table,
with masks drawn uniformly at random, since Proposition~\ref{prop:regimeB}
says they cannot matter. Every trial was compared with the closed form:
for each of the eight bit positions, eight random positions in each of three
$512\times512$ images and one $576\times720$ image, $256$ trials in all, every
one exact. The paper's protocol (ten images, ten positions
each) was then replicated $20{,}000$ times per bit with the closed form to
place the reported mean and spread. The four-round shuffle-then-XOR structure
of our three earlier schemes was modelled with logistic row and column
shuffles and a H\'enon key matrix at Kodak size: a one-bit change alters one
byte, an all-zero chosen image reveals the composed mask, and one
index-encoding chosen image then reveals the composed permutation, whatever
the chaotic parameters.

\paragraph{The UACI check on Jiang et al.}
The per-channel minimum, maximum and average NPCR and UACI of Tables~7 and~8
of~\cite{Jiang2026jisa}, six videos at $512\times512$, were compared with the
independent-ciphertext law at $L=512^2$: NPCR $99.6094\pm0.0122\%$ and UACI
$33.4635\pm0.0462\%$ per frame. The most extreme NPCR entry lies $4.0$
standard deviations from its mean, which is exceeded with probability $0.28$
among $5{,}400$ values (the count if each video had $300$ frames; the paper
does not give it). The UACI extremes lie $19.0$ and $14.6$ standard deviations
away, where the expected largest deviation among $300$ independent frames is $3.0$. The
$\chi^2$ table for encrypted frames (Table~2 of~\cite{Jiang2026jisa}) gives
per-channel averages of $259.2$ to $287.7$; one frame's $\chi^2$ has mean
$255$ and standard deviation $22.6$ at $255$ degrees of freedom.

\paragraph{The surrogate model and the port of Song et al.}
The surrogate model of Algorithm~3 of~\cite{Song2024jisa} flattens the images
of a batch, three bytes per colour pixel; visits positions $j$-major and image
by image among the images that still have a position $j$; and applies $S
\leftarrow k \oplus ((S+k) \bmod 256) \oplus S_{\mathrm{prev}}$, forwards and
then backwards with independent masks. The key-only cross-image permutation is
replaced by a uniform random permutation of all batch positions and the
keystream by uniform bytes. The USC-SIPI batch holds the nine images of the
paper's Table~3, with the least significant bit of Lena at $(130,136)$
flipped; the COCO batch holds the four images of its Table~4, with the least
significant bit of the green channel of image (d) at $(64,80)$ flipped. Over
$1{,}000$ trials per batch the mean NPCR of the listed images had median
$99.890\%$ and $99.808\%$ and fifth percentile $99.633\%$ and $99.626\%$,
against reported means of $99.608\%$ and $99.611\%$.

The port follows the public code (\texttt{Chaos.c},
\texttt{CrossImagePermutation.c}, \texttt{CrossImageDiffusion.c} and
\texttt{main.c}) line by line: the two-dimensional map in single precision
with one pre-iteration and the digit extraction $\lfloor 10^{7}|x|\rfloor$;
the swap-based permutation in the direction \texttt{main.c} uses for
encryption; the keystream map in double precision with $500$ pre-iterations
and $k$ the low byte of the integer part of $10^{15}x$; images visited in identifier order
among those holding position $j$; and seed $123$ at both ends of the chain.
\emph{Equivalence.} The port was checked against the authors' code at commit
\texttt{5f0e2a4} (26 December 2023). \texttt{Chaos.c},
\texttt{CrossImagePermutation.c} and \texttt{CrossImageDiffusion.c} were
compiled unmodified with gcc 13.3 (\texttt{-O0 -std=gnu99 -fno-fast-math
-ffp-contract=off}, x86-64 SSE arithmetic, glibc \texttt{libm}); only the
header of the bitmap loader was replaced by a stand-in with the same image
type and no Windows or file-reading code, and a harness hands the code raw
bytes and writes its intermediates. On a four-image batch of unequal lengths,
batches shaped like the COCO and USC-SIPI batches at reduced size, and the
full-size COCO batch, under the key of \texttt{main.c} and four random keys
(two keys for the full-size batch), the port reproduced the permutation
coordinates, both keystreams, the batch after the forward pass and the
ciphertext byte for byte in all $17$ runs. The keystreams were drawn in the
harness by calling the authors' map function with the byte rule of their
diffusion code; the matching ciphertext confirms the draws made inside the
diffusion functions themselves. The repository carries no licence, so the deposit
fetches the three files by commit hash and does not redistribute them; the
Python port is our own reimplementation for analysis. A compiler or library that rounds the chaotic
maps differently would follow other trajectories, so we cannot confirm that
our runs with their key reproduce theirs.

\emph{Key population.} The code accepts any initial values in $(0,1)$ and
gives no key distribution. Each of $1{,}000$ keys per batch draws the four
initial values uniformly from $(0.05,0.95)$. Keeping away from the ends is a
convention: near $0$ and $1$ the $x(1-x)$ terms of both maps are small, and
we did not want the first iterates to depend on that edge. The results are
conditional on this choice; a sensitivity run with the range widened to
$(0.001,0.999)$ gave counts of $3$ and $13$ for the listed mean and $2$ and $4$
for the largest value (USC-SIPI and COCO, row counted in the displayed
image), against $2$ and $13$, and $2$ and $4$, with the default range. The plaintexts are
scikit-image test images resized to the stated sizes; content enters only
through the sum with the keystream. The released loader stores bitmap rows
bottom first and colour bytes in blue, green, red order, and it allocates the
buffer of an 8-bit image before setting its size, which the port does in the
right order. The stated pixels were placed with the row counted from the top
of the displayed image, which the loader moves to row $H-1-r$, and, as a
sensitivity run, with the row counted from the top of the loaded array; the
green byte is the middle one either way.

\emph{Counts.} With $95\%$ Clopper--Pearson intervals in percent, the listed
mean was at or below the reported one in $17$ of $1{,}000$ surrogate runs
($0.99$--$2.71$) and in $2$ ($0.02$--$0.72$) and $8$ ($0.35$--$1.57$) of
$1{,}000$ port runs for USC-SIPI, and in $12$ ($0.62$--$2.09$), $13$
($0.69$--$2.21$) and $11$ ($0.55$--$1.96$) for COCO; every listed value was
at or below the largest reported one in $3$ ($0.06$--$0.87$), $2$
($0.02$--$0.72$) and $8$ ($0.35$--$1.57$) runs for USC-SIPI and $4$
($0.11$--$1.02$), $4$ ($0.11$--$1.02$) and $1$ ($0.00$--$0.56$) for COCO; COCO
image (d) reached exactly $100\%$ under $996$ ($98.98$--$99.89$) and $999$
($99.44$--$100$) keys. In each triple the order is surrogate, port with the
row counted in the displayed image, port with the row counted in the loaded
array. Over keys, the median of the listed mean was $99.869\%$
(USC-SIPI) and $99.791\%$ (COCO) in the first convention, and the changed pixel
landed on average at $0.43$ and $0.45$ of the chain rather than $0.5$. The
parity constraint of Section~\ref{sec:audit:batch} follows from the update:
the least significant bit of a sum is the XOR of the operands' least
significant bits, so an odd difference at the changed cell leaves the
per-cell difference odd at that cell and at every later cell of the backward
pass, and from the changed cell onwards the accumulated difference is odd
whenever an odd number of cells lie between it and the end of the chain,
inclusive. Cells before the changed one carry the accumulated difference of
the changed cell unchanged.

\section{The detection benchmark}
\label{app:bench}

The simulated ciphers of Table~\ref{tab:bench} act on the flattened image $x$
of length $L$. A round draws a permutation $\pi$ and two byte masks $k,k'$
from a seeded generator, sets $x\leftarrow x_{\pi}$ and applies one forward
and one backward pass. With $\oplus$ for XOR, $+$ for addition modulo $256$
and $S$ for the AES S-box, the forward updates are $c_i=x_i+k_i+c_{i-1}$
(addition), $c_i=x_i\oplus k_i\oplus c_{i-1}$ (XOR),
$c_i=k_i\oplus(x_i+k_i)\oplus c_{i-1}$ (mixed) and $c_i=S[x_i+k_i+c_{i-1}]$
(S-box), with $c_0=0$; the backward pass applies the same update from $i=L$
down with $k'$. The permutation-and-mask cipher is $x_{\pi}+k$ with no
chaining. Two-round ciphers draw fresh $\pi,k,k'$ for the second round. The
hashed variants XOR the seed with the first eight bytes of SHA-256 of the
image, so any change to the image replaces all round material; ``Off'' reruns
them with the plain seed. The null draws both ciphertexts as independent
uniform bytes. Each simulated paper draws a fresh key, five (or ten) of ten
scikit-image test images (camera, moon, brick, grass, gravel, astronaut,
coffee, chelsea, rocket, coins), converted to greyscale or kept in colour and
resized, and one random position per image; the least significant bit and the
top bit are flipped there. The Wu--Noonan--Agaian tests use the mean and
variance of~\cite{Wu2011npcr} with $F=255$ at a per-value level $\alpha'$ with
$1-(1-\alpha')^{m}=0.01$ for $m$ tests per paper. The signature thresholds
are the $1-0.01/(4b)$ quantiles, for $b$ flipped bits, of the four statistics
on $400{,}000$ simulated null tables of binomial NPCR and normal UACI values.
Each threshold depends on one marginal only, so the dependence between NPCR
and UACI of the same ciphertext pair affects only the family rate, and that
rate is measured in the first row of Table~\ref{tab:bench} on papers built
from real byte pairs.

\paragraph{Recalibrated thresholds of Table~\ref{tab:ablation}}
Each test of Table~\ref{tab:ablation} is written as one or more table-level
statistics, each oriented so that larger is more extreme: for the per-value
Wu tests, the largest deviation of the five values from the reference (one-
or two-sided, NPCR or UACI); for the table-level rules, the minimum, the
absolute deviation of the mean, and the NPCR and UACI standard deviations.
A test flags a table when any of its $s$
statistics exceeds its $1-\alpha$ quantile on the calibration sample, with
one common level $\alpha$ for all $s$ statistics. The level is found by
bisection ($40$ steps on $[0.01/(2s),\,0.01]$), keeping the largest level
found whose rate on the calibration sample stays below $1\%$. Ties at a threshold count as not flagged, so for the discrete,
rounded statistics the achieved rate can sit a little below $1\%$. The
calibration and validation samples are two independent sets of $400{,}000$
tables from the exact joint null; the ``Recal.'' column of
Table~\ref{tab:ablation} is the rate on the validation sample.

\begin{table*}[!t]\centering
\caption{The benchmark of Table~\ref{tab:bench} under other reporting conditions: share of
papers flagged (\%), five trials per paper, $1{,}000$ papers per cell. W:
Wu--Noonan--Agaian NPCR and UACI tests, least significant bit. S: signature
check, least significant bit. O: signature check on two bits with
plaintext-derived material switched off. ``dp'' is decimal places printed.}
\label{tab:benchrobust}\small\setlength{\tabcolsep}{3.5pt}
\begin{tabular}{@{}l*{12}{r}@{}}\toprule
 & \multicolumn{3}{c}{$256^2$, 4 dp} & \multicolumn{3}{c}{$256^2$, 2 dp} & \multicolumn{3}{c}{$512^2$, 4 dp} & \multicolumn{3}{c}{$256^2\times3$, 4 dp}\\
\cmidrule(lr){2-4}\cmidrule(lr){5-7}\cmidrule(lr){8-10}\cmidrule(lr){11-13}
Cipher & W & S & O & W & S & O & W & S & O & W & S & O\\ \midrule
Independent (null) & 1.0 & 0.4 & 0.5 & 1.1 & 0.5 & 0.4 & 0.8 & 1.2 & 1.4 & 1.1 & 0.5 & 0.6\\
Permutation, mask & 100.0 & 100.0 & 100.0 & 100.0 & 100.0 & 100.0 & 100.0 & 100.0 & 100.0 & 100.0 & 100.0 & 100.0\\
One round, addition & 100.0 & 100.0 & 100.0 & 100.0 & 100.0 & 100.0 & 100.0 & 100.0 & 100.0 & 100.0 & 100.0 & 100.0\\
One round, XOR & 100.0 & 100.0 & 100.0 & 100.0 & 100.0 & 100.0 & 100.0 & 100.0 & 100.0 & 100.0 & 100.0 & 100.0\\
One round, add./XOR & 100.0 & 100.0 & 100.0 & 100.0 & 100.0 & 100.0 & 100.0 & 100.0 & 100.0 & 100.0 & 100.0 & 100.0\\
One round, S-box & 6.4 & 100.0 & 100.0 & 6.9 & 100.0 & 100.0 & 34.3 & 100.0 & 100.0 & 23.4 & 100.0 & 100.0\\
Two rounds, addition & 1.6 & 2.1 & 100.0 & 1.7 & 1.5 & 100.0 & 2.5 & 2.0 & 100.0 & 2.0 & 2.0 & 100.0\\
Two rounds, S-box & 1.3 & 0.8 & 0.4 & 1.3 & 0.8 & 0.3 & 1.1 & 2.0 & 1.6 & 1.4 & 0.7 & 0.6\\
A, hashed key & 1.1 & 1.1 & 100.0 & 1.1 & 1.0 & 100.0 & 2.0 & 1.4 & 100.0 & 1.5 & 1.4 & 100.0\\
B-add, hashed key & 1.0 & 0.4 & 100.0 & 1.3 & 0.2 & 100.0 & 1.1 & 1.1 & 100.0 & 1.0 & 0.5 & 100.0\\
\bottomrule\end{tabular}\end{table*}

\section{The checker}
\label{app:checker}

\texttt{npcr\_check} reads a list of tables. Each entry gives the values, or
their mean, standard deviation and count; the smallest and largest number of
bytes $L$ behind them; the trials per value (one, a stated number, or
unknown); the regime read from the equations; whether the key or nonce
depends on the plaintext; the printed precision (\texttt{decimals}, and
\texttt{sd\_decimals} for a printed summary SD, which defaults to
\texttt{decimals}); and, optionally, the UACI extremes. The rules below are
applied in this order, and the first verdict reached is kept; later rules
still add their flags.

(i) A plain-versus-cipher table is not a differential measurement. (ii) A
fixed-material regime A structure cannot give a mean above its structural
bound, $100r/L$ for $r$ affected bytes, plus half a rounding step. (iii)
Plaintext-dependent material is flagged, with the advice to rerun with the
dependence off. (iv) A mean below $1\%$ is reported as limited propagation
(only the equations can show that it is the no-propagation structure of
regime A), and a
mean near $50\%$ with a standard deviation above ten points as compatible
with XOR chaining or a top-bit change. (v) A value below $99\%$ is reported
as a departure from the reference. For a least-significant-bit change under
additive chaining ($q=256$) such a value requires a cancellation, but for
other $q$ it need not (with $q=2$, $L=65{,}536$ and $m=32{,}769$ the count
gives $75\%$ with no cancellation), and outside the model other causes are
possible. A table is labelled two clusters when at least three
values lie below $99\%$ with standard deviation under one point, at least
three lie at or above $99\%$, and the two group means differ by more than ten
points, since regime B cancellations spread over the whole interval. (vi) For
single trials the spread is tested in both tails, at $0.5\%$ each, against
the sample standard deviations of $200{,}000$ simulated null tables of $n$
binomial values rounded to the printed precision: at the smallest $L$ for the
upper tail and at the largest $L$ for the lower, with ties counted against
the table. A printed summary SD is compared at the edge of its own rounding
interval. The Monte Carlo $p$-value is $(c+1)/(M+1)$ for $c$ exceedances in
$M$ simulated tables. When no simulated table reaches an observed spread, we
report only that none of the $M=200{,}000$ did, with the $95\%$ upper bound
$3/M=1.5\times10^{-5}$ on the tail probability. We give no smaller value: the
chi-square form with a rounding variance $h^2/12$ added, for rounding step
$h$, is an approximation and does not bound the tail of the rounded binomial
model. When the precision is neither stated nor inferable from the values,
the lower tail is withheld. An upper-tail excess is labelled
compatible with a one-round chaining signature when the mean lies between
$99\%$ and $100\%$ and either the mean is within $0.15$ points of the
no-cancellation value $99.805\%$ or the spread exceeds $0.05$ points;
otherwise it is reported as a departure from regime C with the cause
unresolved. A lower-tail result is
labelled a spread too small for independent single trials. (vii) The mean is
tested with standard error $\sqrt{(\sigma_C^2/k+h^2/12)/n}$ for $k$ trials per
value, where $h^2/12$ approximates the variance that rounding adds by treating
the rounding error as uniform and independent of the value, and a systematic excess with a two-sided sign test. Its probability per
value is the exact chance that a printed value lies above the reference,
$P\{\mathrm{Bin}(kL,255/256)>x_h\}$, where $x_h$ is the smallest count that
prints above $99.609375\%$ at the stated precision (for mixed sizes the
larger $p$ over the extreme sizes is used). When $k$ is not stated no
probability is given and only the count of values above is reported. The test
assumes independent values and is reported as a flag. (viii) UACI extremes beyond
five standard deviations of the independent-ciphertext law are flagged; this
is a screening rule, not a calibrated test, because the number of values
behind the extremes is usually not stated.

Tests (vi) and (vii) run at the $1\%$ level. Averages, single values and one
trial per batch are reported as undecidable from the table where the spread
cannot be read, with the reason. When no test rejects, the verdict is ``no
discrepancy with regime C'': the implemented tests found nothing, which is
weaker than a certificate that the values are independent. Each verdict
carries a basis: the protocol, the printed equations, or the table alone, and
in the last case a label that names a structure is marked as a statistical
suggestion. Printed precision is taken from the input or, failing that, from
the number of decimals in the values, which a dropped trailing zero can only
make coarser and the test more conservative.

\paragraph{Validation}
The validation of Section~\ref{sec:protocol:tool} fed each simulated paper of
Appendix~\ref{app:bench} to the checker in this form, with nothing about the
structure and, for the hashed variants, the statement that the key depends on
the image. The null and alternative tables of Table~\ref{tab:checkval} come
from a random stream separate from the one that builds the checker's
reference tables, so calibration and validation share no draws. Every table
has $n=5$ values, rounded to the stated precision and entered as single trials
with that precision declared. In the null tables each NPCR and UACI pair comes
from one simulated pair of independent uniform ciphertexts of $L$ bytes: the
counts of the byte differences $d=|X-Y|$ are multinomial, with $P(d=0)=1/256$
and $P(d=k)=2(256-k)/256^2$, so NPCR $=100(1-n_0/L)$ and UACI
$=100\sum_k k\,n_k/(255L)$ keep their joint law. They are not independent: for
one byte the indicator of a change and $|X-Y|/255$ have correlation $0.089$.
The checker receives the five NPCR values and the smallest and largest UACI
value. The alternatives
use $L=65{,}536$ and four decimals, give the checker NPCR values only, and are
generated directly as NPCR values rather than from ciphertext bytes: (a) each
value is the mean of ten independent binomial values as above; (b) each value
comes, independently and with probability $1/2$, from regime B (the count of
Proposition~\ref{prop:regimeB} for a least-significant-bit change under
addition, $j$ uniform on $1,\dots,L$) and otherwise from regime C; (c)
$v_i = 100\cdot255/256 + \sigma_C\bigl(\sqrt{\rho}\,Z_0 + \sqrt{1-\rho}\,Z_i\bigr)$
with $\rho=0.9$ and $Z_0,\dots,Z_5$ independent standard normal, a Gaussian
model of values that share randomness; (d) each value is a binomial count with
mismatch probability $1-1.2/256$, which is the NPCR law of two independent
ciphertexts whose bytes have collision probability $1.2/256$, although only
that marginal is modelled, not the bytes or the UACI; (e) each value is the
mean of ten regime B values drawn as in (b).

\begin{table*}[!t]\centering
\caption{Counts behind the checker validation of Section~\ref{sec:protocol:tool}.
Flagged: tables not returned as ``no discrepancy with regime C''. Interval:
$95\%$ Clopper--Pearson for the flagged share, in percent. Lower: share
flagged by the lower-tail spread test. Chain: share labelled compatible with a
one-round chaining signature. Alternatives (a)--(e) are defined in the text.}
\label{tab:checkval}\small
\begin{tabular}{@{}lrrrrr@{}}\toprule
Condition & Tables & Flagged & Interval & Lower & Chain\\ \midrule
Null, $256^2$, 4 dp & 100{,}000 & 2{,}056 & 1.97--2.15 & 0.48 & 0.24\\
Null, $256^2$, 2 dp & 50{,}000 & 937 & 1.76--2.00 & 0.32 & 0.25\\
Null, $512^2$, 4 dp & 50{,}000 & 972 & 1.82--2.07 & 0.54 & 0.00\\
Null, $512^2$, 2 dp & 50{,}000 & 1{,}159 & 2.19--2.45 & 0.43 & 0.00\\
Null, $1024^2$, 4 dp & 50{,}000 & 999 & 1.88--2.12 & 0.49 & 0.00\\
Null, $1024^2$, 2 dp & 50{,}000 & 523 & 0.96--1.14 & 0.00 & 0.00\\
\midrule
(a) Averages of ten trials as singles & 5{,}000 & 1{,}392 & 26.60--29.11 & 27.84 & 0.00\\
(b) B/C mixture, $p=1/2$ per value & 5{,}000 & 4{,}601 & 91.23--92.76 & 0.02 & 88.64\\
(c) Common component, $\rho=0.9$ & 5{,}000 & 2{,}252 & 43.65--46.43 & 28.20 & 0.00\\
(d) Mismatch probability $1-1.2/256$ & 5{,}000 & 5{,}000 & 99.93--100.00 & 0.34 & 0.54\\
(e) Regime B averaged over ten & 5{,}000 & 5{,}000 & 99.93--100.00 & 0.08 & 18.12\\
\bottomrule\end{tabular}\end{table*}

\begin{table*}[!t]\centering
\caption{Labels returned by \texttt{npcr\_check} against the simulated mechanism,
least significant bit, share of tables in percent ($1{,}000$ papers per
benchmark cipher, $5{,}000$ tables per alternative). None: no discrepancy with
regime C. Lim.: limited propagation. Chain: compatible with a one-round chaining
signature. XOR: compatible with XOR chaining or a top-bit change. Small: spread
too small for independent single trials. Clust.: two clusters. Unres.:
departure from regime C, cause unresolved. Entries off the expected label,
such as chaining suggested for non-uniform bytes, are the wrong suggestions.
Five-value tables cannot receive the two-cluster label; that rule is tested
in Table~\ref{tab:clusters}.}
\label{tab:confusion}\small
\begin{tabular}{@{}lrrrrrrr@{}}\toprule
Simulated mechanism & None & Lim. & Chain & XOR & Small & Clust. & Unres.\\ \midrule
Independent ciphertexts (null) & 98.4 & 0.0 & 0.1 & 0.0 & 0.3 & 0.0 & 1.2\\
Permutation and mask & 0.0 & 100.0 & 0.0 & 0.0 & 0.0 & 0.0 & 0.0\\
One round, addition & 0.0 & 0.0 & 94.4 & 0.0 & 0.0 & 0.0 & 5.6\\
One round, XOR & 0.0 & 0.0 & 0.0 & 53.7 & 0.0 & 0.0 & 46.3\\
One round, add./XOR & 0.0 & 0.0 & 96.8 & 0.0 & 0.0 & 0.0 & 3.2\\
One round, S-box & 0.0 & 0.0 & 95.2 & 0.0 & 0.0 & 0.0 & 4.8\\
Two rounds, addition & 96.0 & 0.0 & 0.8 & 0.0 & 0.4 & 0.0 & 2.8\\
Two rounds, S-box & 98.6 & 0.0 & 0.0 & 0.0 & 0.3 & 0.0 & 1.1\\
A, hashed key & 98.0 & 0.0 & 0.3 & 0.0 & 0.4 & 0.0 & 1.3\\
B-add, hashed key & 98.0 & 0.0 & 0.0 & 0.0 & 0.8 & 0.0 & 1.2\\
\midrule
(a) Averages as singles & 72.2 & 0.0 & 0.0 & 0.0 & 27.8 & 0.0 & 0.0\\
(b) B/C mixture & 8.0 & 0.0 & 88.6 & 0.0 & 0.0 & 0.0 & 3.4\\
(c) Common component & 55.0 & 0.0 & 0.0 & 0.0 & 28.2 & 0.0 & 16.8\\
(d) Non-uniform bytes & 0.0 & 0.0 & 0.5 & 0.0 & 0.3 & 0.0 & 99.1\\
(e) Regime B averaged & 0.0 & 0.0 & 18.1 & 0.0 & 0.1 & 0.0 & 81.8\\
\bottomrule\end{tabular}\end{table*}

\begin{table*}[!t]\centering
\caption{The two-cluster rule on tables of $n$ values, $5{,}000$ tables per
condition, $L=65{,}536$, four decimals; the conditions are defined in the text. Elig.: share of tables with at least
three values on each side of $99\%$. Label: share labelled two clusters, with
its $95\%$ Clopper--Pearson interval, in percent.}
\label{tab:clusters}\small
\begin{tabular}{@{}lrrrrrr@{}}\toprule
& \multicolumn{3}{c}{$n=10$} & \multicolumn{3}{c}{$n=28$}\\
\cmidrule(lr){2-4}\cmidrule(lr){5-7}
Condition & Elig. & Label & Interval & Elig. & Label & Interval\\ \midrule
Two clusters & 73.8 & 73.8 & 72.5--75.0 & 99.9 & 99.9 & 99.8--100.0\\
Regime B, addition & 0.0 & 0.0 & 0.0--0.1 & 0.0 & 0.0 & 0.0--0.1\\
Regime B, XOR & 0.1 & 0.0 & 0.0--0.1 & 0.3 & 0.0 & 0.0--0.1\\
B/C mixture & 0.0 & 0.0 & 0.0--0.1 & 0.0 & 0.0 & 0.0--0.1\\
Independent (null) & 0.0 & 0.0 & 0.0--0.1 & 0.0 & 0.0 & 0.0--0.1\\
\bottomrule\end{tabular}\end{table*}

\paragraph{The two-cluster rule}
The rule needs at least three values on each side of $99\%$, so the
five-value tables above cannot trigger it, and the zero column ``Clust.'' of
Table~\ref{tab:confusion} says nothing about it. We therefore tested it
separately on $5{,}000$ tables each of $n=10$ and $n=28$ values
($L=65{,}536$, four decimals, single trials, from a separate random stream):
a two-cluster alternative in which each value, independently with probability
$10/28$, comes from a change that reaches half the image,
$100\,\mathrm{Bin}(L/2,255/256)/L\approx49.8\%$, and otherwise from regime C,
as the design of Section~\ref{sec:audit:liao} would give; two broad unimodal
alternatives, regime B with addition ($q=256$) and with XOR ($q=2$), $j$
uniform; the B/C mixture of alternative (b); and null tables from the exact
joint law. Table~\ref{tab:clusters} gives the results. The label was never
given outside the two-cluster alternative, including the regime B XOR tables
that have three or more values on each side, whose low values spread too
widely. Within it, the label was given whenever at least three values fell
on each side, so its sensitivity at $n=10$, $74\%$, is set by that
requirement, and at $n=28$ it is $99.9\%$. These are the cases the rule was
written for; a mixture of two unimodal laws closer together than ten points,
or with a low cluster wider than one point, would not receive it.

\section{Finite-state results used in Section~\ref{sec:eval:graph}}
\label{app:finite}

The map of Section~\ref{sec:eval:graph} acts on
$S_w=(\mathbb{Z}/2^w\mathbb{Z})^n$. With $\mathrm{MULHI}(u,v)=\lfloor
uv/2^w\rfloor$, $t_{i-1}=\mathrm{MULHI}(X_{i-1},\,-X_{i-1}\bmod 2^w)$ and
$Q(X_{i-1})=\mathrm{MULHI}(t_{i-1},t_{i-1})$, one synchronous update is
$T(X)_i = 3X_i + X_{i-1} + \varepsilon\,Q(X_{i-1}) \bmod 2^w$, indices modulo
$n$. Its linear core is $X\mapsto AX \bmod 2^w$ with $A=aI+S$, $a=3$ and $S$
the cyclic shift. The results below are proved in full in the deposited
report; we give the statements and the short arguments.

\emph{Determinant and invariant factors.} $A$ is the circulant with symbol
$a+z$, so $\det A=\prod_{\omega^n=1}(a+\omega)=a^n-(-1)^n$. In
$\mathbb{Z}^n/A\mathbb{Z}^n$ each relation $Ae_i\equiv0$ reads
$Se_i\equiv-a\,e_i$, so every basis vector is a multiple of the class of
$e_1$ and the quotient is cyclic of order $d=\lvert\det A\rvert$. The Smith
normal form of $A$ is therefore $\operatorname{diag}(1,\dots,1,d)$.

\emph{Image of the linear core.} With $UAV=\operatorname{diag}(1,\dots,1,d)$
for unimodular $U,V$, which stay invertible modulo $2^w$, the image of the
core has the size of the image of $\operatorname{diag}(1,\dots,1,d)$, whose
last coordinate has index $\gcd(d,2^w)=2^{\min(w,k)}$, $k=v_2(\det A)$. The
image fraction is $2^{-\min(w,k)}$. At $a=3$, $\det A=8$ ($k=3$) for $n=2$ and
$80$ ($k=4$) for $n=4$, so the fraction is $1/8$ and $1/16$ for every $w\ge k$.
For a general integer matrix the fraction is $2^{-\sum_i\min(w,v_2(d_i))}$
over its invariant factors $d_i$; the formula above uses the fact that here
only one of them exceeds one.

\emph{Collapse of the linear core.} For $a$ odd, $A\equiv I+S \pmod 2$, and
for $n$ a power of two $(I+S)^n=\sum_j\binom{n}{j}S^j\equiv I+S^n=0 \pmod 2$.
So $A^n=2B$ for an integer matrix $B$, $A^{nw}\equiv 0 \pmod{2^w}$, every
state reaches $0$ within $nw$ steps, and $0$ is the only cyclic state. For
$n=3$ this fails: at $a=3$, $w=1$ the states $(1,1,0)$, $(1,0,1)$, $(0,1,1)$
form a cycle.

\emph{Parity bound for the full map.} For $a$ odd and $\varepsilon$ even,
summing the coordinates of $T(X)$ gives $(a+1)\sum_iX_i+\varepsilon\sum_i
Q(X_i)\equiv0 \pmod 2$, so every successor lies in the half
$H=\{Y:\sum_iY_i\ \text{even}\}$ and the image fraction is at most $1/2$. If
$n$ is also a power of two, then modulo $2$ the map is $X\mapsto(I+S)X$, so
$T^n(X)$ has every coordinate even and all cyclic states lie in a set of
$2^{n(w-1)}$ states; at $n=4$, $w=32$ that is $2^{124}$ of $2^{128}$. This is
a bound on the state transition, not a statement about the extracted words.

\emph{Enumeration.} Table~\ref{tab:finite} gives the exhaustive values for six
cells, every state evaluated. The core column agrees with the formula above.
For $\varepsilon=2$ the relevant reference is a random map from $S_w$ into
$H$, whose image fraction is $(1-e^{-2})/2=0.432$; the unrestricted
random-mapping value $1-e^{-1}=0.632$ applies to odd $\varepsilon$. The
reference value $\varepsilon=2$ was selected, before the parity bound was
known, by a rule fixed in advance that scored candidates by closeness to
$1-e^{-1}$ on image fraction and cyclic-node count; the bound shows that
even values could not meet that criterion.

\begin{table}[!t]\centering
\caption{Exhaustive enumeration of the map at reduced sizes. Image fraction:
share of states that are the image of some state. Cyclic: number of states on
cycles. Core: the linear core. The random-map reference for $\varepsilon=1$
is $1-e^{-1}=0.632$; for $\varepsilon=2$, confined to the even-parity half,
it is $(1-e^{-2})/2=0.432$.}
\label{tab:finite}\small\setlength{\tabcolsep}{3.5pt}
\begin{tabular}{@{}lrrrrrr@{}}\toprule
& \multicolumn{3}{c}{Image fraction} & \multicolumn{3}{c}{Cyclic states}\\
\cmidrule(lr){2-4}\cmidrule(lr){5-7}
$(n,w)$ & core & $\varepsilon=1$ & $\varepsilon=2$ & core & $\varepsilon=1$ & $\varepsilon=2$\\ \midrule
$(2,8)$  & 0.1250 & 0.6539 & 0.4527 & 1 & 456 & 313\\
$(4,4)$  & 0.0625 & 0.2897 & 0.2057 & 1 & 1 & 23\\
$(2,10)$ & 0.1250 & 0.6554 & 0.4477 & 1 & 868 & 733\\
$(4,5)$  & 0.0625 & 0.6978 & 0.4604 & 1 & 498 & 222\\
$(2,11)$ & 0.1250 & 0.6540 & 0.4497 & 1 & 4{,}614 & 1{,}441\\
$(4,6)$  & 0.0625 & 0.6634 & 0.4403 & 1 & 13{,}121 & 2{,}036\\
\bottomrule\end{tabular}\end{table}

\FloatBarrier

\bibliographystyle{cas-model2-names}
\bibliography{references}

\end{document}